# Antisymmetry: Fundamentals and Applications

Hari Padmanabhan,[1,*] Jason M. Munro,[1,2,*] Ismaila Dabo,[1,2] Venkatraman Gopalan,[1,3,4]

[1] Department of Materials Science and Engineering, and Materials Research Institute, The Pennsylvania State University, University Park, PA, 16802, USA; vgopalan@psu.edu, hari@psu.edu

[2] Penn State Institutes of Energy and the Environment, The Pennsylvania State University, University Park, PA 16802, USA; dabo@psu.edu, Jason.munro@gmail.com

[3] Department of Physics, The Pennsylvania State University, University Park, PA, 16802, USA;

[4] Department of Engineering Science and Mechanics, The Pennsylvania State University, University Park, PA, 16802, USA.

* equal contributions.

Abstract

Symmetry is fundamental to understanding our physical world. An antisymmetry operation switches between two different states of a trait, such as two time-states, position-states, charge-states, spin-states, chemical-species etc. This review covers the fundamental concepts of antisymmetry, and focuses on four antisymmetries, namely spatial inversion in point groups, time reversal, distortion reversal and wedge reversal. The distinction between classical and quantum mechanical descriptions of time reversal is presented. Applications of these antisymmetries in crystallography, diffraction, determining the form of property tensors, classifying distortion pathways in transition state theory, finding minimum energy pathways, diffusion, magnetic structures and properties, ferroelectric and multiferroic switching, classifying physical properties in arbitrary dimensions, and antisymmetry-protected topological phenomena are presented.

# 1. INTRODUCTION AND SCOPE

Symmetry is *doing something that looks like doing nothing*. This statement turns out to be a pretty rigorous basis for the mathematical definition of a symmetry operation. Symmetry lies at the heart of understanding the physical world, from fundamental laws of physics (1), to materials and their physical properties (2).Any introductory book on materials science (3) begins with a description of crystal structure classified according to crystallographic symmetries. While such symmetry classification dates back to the 1891 work of Fedorov (4) and Schoenflies (5, 6), the discovery in 1912 by Friedrich, Knipping & Laue (7) that crystals are periodic array of atoms laid the foundation stone for the modern materials science and condensed matter physics.

Antisymmetry is a specific type of symmetry, also called 'two-color' symmetry. The reader is referred to several excellent reviews on the topic from decades earlier (8, 9). For example, spatial inversion, $\bar{1}$, is an antisymmetry with respect to rotations. Time reversal, $1'$, is a well-known antisymmetry with applications in magnetic crystallography and in predicting magnetic properties with classical spins. A new antisymmetry, distortion reversal, $1*$, has recently been introduced in predicting minimum energy pathways (10). Another recently formulated antisymmetry, wedge reversion (11), $1^{\dagger}$, enables the classification of physical quantities and properties in arbitrary dimensions. This review aims to overview the fundamental concepts underlying antisymmetry, and the specific examples of antisymmetries mentioned above, namely, $\bar{1}$, $1'$, $1*$ and $1^{\dagger}$.

Section 2 introduces the concepts of symmetry, color symmetry and antisymmetry. Section 3 discusses time reversal antisymmetry, $1'$, its applications to magnetic crystallography and property predictions, and its interpretation in classical versus quantum mechanical contexts. Section 4 discusses distortion reversal antisymmetry, $1*$, and its applications in finding minimum energy pathways. Section 5 introduces a new antisymmetry, wedge reversion, $1^\dagger$ and its application in classifying multivectors in arbitrary dimensions. Section 6 presents a summary and future outlook including a brief note on topology. We restrict ourselves to crystallographic symmetry. All the physical problems addressed here are in the non-relativistic Newtonian framework, where space is considered 3-dimensional, and time is considered a scalar quantity disconnected from space. Relativistic 4D spacetime involves three fundamental symmetries of antiparticle conjugation ($C$), parity ($P$) and time reversal ($T$) (12). Parity and time reversal are discussed in the non-relativistic context as spatial inversion, $\bar{1}$, and time-reversal, $1'$ respectively. Antiparticle conjugation is related to four-dimensional (4D) spacetime reversal (inverting both space and time coordinates), and is not discussed here further, other than in the non-relativistic context as $\bar{1}'$.

## 2. SYMMETRY AND ANTISYMMETRY

## 2.1. Crystallographic Point and Space groups, Spatial Inversion, $\bar{1}$ and Neumann's Principle

*Rotations* are the most common symmetry operations in crystallography. It turns out from simple geometric considerations (see **Figure 1a**) that periodic crystals can only possess five types of rotational symmetries, namely 1-, 2-, 3-, 4- and 6- fold rotations, where a *p*-fold axis refers to rotations of $\frac{2\pi}{p}$ about that axis. This is called the crystallographic restriction theorem and is a direct consequence of the translational periodicity of crystals. If we now consider a periodic crystal, but for the moment ignore translational symmetry, it can be shown that there are only 11 possible ways in which a collection of the above rotational symmetries about different axes can pass through a single point in a self-consistent manner (see **Figure 1b**). These are the 11 rotational *point groups*, listed in the **Figure 1c**. The point group 32 is shown as an example in the panel **b** of **Figure 1**.

Spatial inversion, $\bar{1}$, inverts the space about a single point of inversion, namely, $\bar{1}: \boldsymbol{r} \rightarrow -\boldsymbol{r}$. In two dimensions (2D), $\bar{1}$ is equivalent to a 2-fold rotation of the 2D plane since both can be represented by $(x\,y) \rightarrow (-x-y)$. In three dimensions (3D), they are distinct since $\bar{1}: (x\,y\,z) \rightarrow (-x-y-z)$ while the 2-fold transformation along $z$-axis is $2_z: (x\,y\,z) \rightarrow (-x-y\,z)$. Rotoinversions combine rotation $R$ with inversion, $\bar{1}$ to form the operation, $R \cdot \bar{1} = \bar{R}$. These are distinct operations from $R$ and $\bar{1}$ since an object can possess $\bar{R}$ symmetry without possessing either $R$ or $\bar{1}$. Rotations and rotoinversions of a periodic crystal that leave the crystal invariant together compose the 32 crystallographic point groups (13). When three dimensional translations in the Euclidean space are also included, the 32 point-groups expand to 230 space-groups as listed in detail in the International Table of Crystallography (14).

Neumann's principle states that the macroscopic properties of a material must *at least* possess the symmetry of its point group, thus tying symmetry and properties together intimately (2). By macroscopic, it is meant that the property in question is averaged over sufficiently large number of unit cells for it to be insensitive to lattice translation symmetry. For example, if the point group of a crystal possesses inversion symmetry, $\bar{1}$, then all its macroscopic properties must at least possess inversion symmetry, thus ruling out macroscopic polarization such as ferroelectricity and pyroelectricity within this crystal. When microscopic properties that involve unit cell translations, Neumann's principle needs to be restated as follows: microscopic properties involving lattice translations of a material must *at least* possess the symmetry of its space group. For example, the arrangement of spins on different lattice sites related by translations determines the antiferromagnetic order (15, 16). Inversion symmetry is also useful in quantum mechanics; for example, one of the optical selection rules (17) states that dipole transition of electrons through light field can only occur between two electronic states with opposite parities, namely one state being $\bar{1}$-even (invariant under $\bar{1}$, such as $s$, $d$, $g$ orbitals) and the other being $\bar{1}$-odd (reverses under $\bar{1}$, e.g. $p$, $f$, $h$ orbitals).

## 2.2. Color Symmetry and Antisymmetry

Symmetry operations in the crystallographic point groups and space groups exclusively involve moving atoms around in the 3D Euclidean space, i.e. the spatial location of atoms. However, atoms also can possess other characteristics, for example, the type of chemical species, charge and spin. These characteristics can be thought of as "color" in an abstract sense. If there are only two states of charge, say + or -, or say only two states of quantum spin, namely, ↑ and ↓, then we need only two colors to represent them, say white for ↑ and black for ↓. Two-color symmetries are called antisymmetries or anti-identities; an *antisymmetry* operation switches a black characteristic into a white characteristic and vice versa. **Figure 2** shows an example of a two-color and a three-color symmetry. Review articles on color symmetry by Lifshitz (18), Schwarzenberger (19), and Opechowski (20) are particularly recommended. This review will solely focus on two-color symmetries, also called antisymmetries. Three or more color permutation groups are distinct from the antisymmetry groups to be discussed in this review.

An antisymmetry with respect to a group can be defined (22) as an operation satisfying the following conditions - (1) it should be its own inverse (i.e. self-inverse), (2) it must commute with all elements of the group under consideration, and (3) it must not itself be an element of the group. By this definition, the spatial inversion operation is in fact an antisymmetry with respect to the group of proper rotations, $SO(3)$. However, note that inversion is not an antisymmetry with respect to the Euclidean group $E(3)$, since it does not commute with spatial translations.

A brief note on the history of antisymmetry is in order, e. g. see articles by Zamorzaev (9) and Wills (21), and the bibliography within them. Heesch introduced the idea of antisymmetry in 1930 (22) and showed that color could be treated as a higher dimension. By including a single antisymmetry to the crystallographic groups, Heesch showed that they could be expanded into 122 point groups, now called Shubnikov groups. Zamorzaev (23, 24) extended these two-color point groups to 2 color space groups, expanding the 230 crystallographic space groups to 1651 two-color space groups. Landau and Lifshitz (25) reinterpreted black and white colors to correspond to time reversal antisymmetry, $1'$ whose action reverses time, i.e. $t \rightarrow$ -$t$. These groups revolutionized the interpretation of neutron scattering from magnetic crystals, and in interpreting magnetic properties of crystals as long as the spins are treated classically. However, there are subtle conceptual issues in conflating time reversal with spin reversal, especially in quantum mechanics, which is discussed in Section 3.2 below. Gopalan and Litvin introduced (26) an antisymmetry named *rotation reversal symmetry* in 2011 as a way to switch the sense of solid-body rotations of rigid polyhedra that compose a crystal. This idea was generalized to *distortion reversal antisymmetry*, $1^*$, by VanLeeuwen and Gopalan, in 2015 (10, 27). It is shown in Section 4 to have important practical applications in predicting minimum energy pathways between an initial and a final state (10, 28, 29). Gopalan introduced wedge reversion, $1^{\dagger}$, in 2019 as a missing antisymmetry in the classification of physical properties expressed as multivectors (11) into 8 principal and 41 overall types.

## 2.3. Antisymmetry Point Groups and Space Groups

Starting from 32 crystallographic point groups and 230 space groups, addition of antisymmetry operations expands the groups further. Consider including a single antisymmetry to the crystallographic groups, resulting in the so called single antisymmetry point groups (SAPG) and single antisymmetry space groups (SASG). The most common application of these groups is in the form of 122 Shubnikov point groups and 1651 space groups, all listed by Litvin (30, 31), where the antisymmetry is $1'$, the time-reversal operation. Of the 122 two-color point groups, 32 of them are *colorless groups*, the ones with no $1'$, and thus they are the conventional 32 crystallographic groups. Another 32 of them have $1'$ explicitly in them, and they are called *grey groups* -these groups cannot support a macroscopic magnetic moment, since Neumann's principle would require that the presence of a magnetic moment, $\boldsymbol{M}$, will also require the presence of $1'\boldsymbol{M} = -\boldsymbol{M}$ in the system. Finally, there are 58 more where $1'$ is not present explicitly, but is present in association with other rotation or rotoinversion operations, such as in e.g. $2'=2. 1'$. These are called *black and white groups*. The colorless point groups (32 of them) and the black & white point groups (58 of them) can support a macroscopic magnetic moment, and hence together, they are called the 90 magnetic groups. See **Figure 3** for examples of colorless, grey, and black and white SAPGs.

If two antisymmetry operations are defined and represented as 1* and $1'$, they can be used to generate double antisymmetry point groups (DAPGs) and double antisymmetry space groups (DASGs). In all, 17,803 possible DASGs and 624 DAPGs exist, as listed by VanLeeuwen, Gopalan and Litvin (32). Huang, VanLeeuwen, Litvin and Gopalan (33) published the complete symmetry diagrams for all of these groups. Padmanabhan et al listed all spatio-temporal point groups with time translations (34), and Liu et al listed all spatio-temporal point groups with time translations and time reversal (35). If time reversal is replaced by 1*, these listings would also apply to isomorphic distortion groups described later in Section 4. **Figure 4** shows example stereographic projections of DAPGs. The addition of a third antisymmetry leads to 287,574 triple antisymmetry space groups (TASGs) (27) and 4,362 point groups (TAPGs) (9); these are not yet explicitly listed. Beyond 3 antisymmetries, one can only construct grey point groups that will explicitly contain the additional antisymmetry operations. This is because of a lack of greater than 3 distinct subgroups of index 2 (containing half the number of elements) of the original crystallographic point groups needed to construct black and white point groups.

## 3. TIME REVERSAL AS AN ANTISYMMETRY

Landau and Lifshitz (36) reinterpreted the generalized antisymmetry operation $1'$ as a time-reversal operation defined as follows: $1': t \rightarrow -t$. Classically, magnetic moments can be thought of as current loops, and hence a time-reversal operation will flip the direction of the current in the loop and hence the magnetic moment. Magnetic space and point groups, along with the advent of neutron diffraction, transformed the way magnetic materials and their properties were characterized. However, the above description of time-reversal as an antisymmetry, and its action on magnetic moments is only valid within a purely classical treatment of spin. The intrinsic spins of electrons are inherently quantum mechanical in origin, and behave very differently from classical current loops, and care is needed in applying magnetic groups in these cases, as described below.

### 3.1. Neutron Diffraction Reveals Magnetic Symmetry

The antisymmetry operation $1'$ is used to reverse classical magnetic moments (equivalent to a flat current loop) between up ($\uparrow$) and down ($\downarrow$) states at each magnetic atom, while simultaneously leaving their spatial coordinates unchanged. The data in **Figure 5a** is the first instance of the application of magnetic groups to neutron diffraction of a magnetic crystal(37, 38). Conventional X-ray photons do not distinguish between spin up and spin down electrons; they simply see the total electron density, $\rho_e(r)$, and so their scattering power $f_\gamma \propto \rho_e(r) = \rho_e^\uparrow(r) + \rho_e^\downarrow(r)$, where the superscripts indicate the two spin states. Neutrons scatter due to nuclear spin as well as the magnetic polarization of the electrons, namely, $f_{n^0}^{mag} \propto \rho_e^\uparrow(r) - \rho_e^\downarrow(r)$, and hence can probe the underlying magnetic symmetry of the crystal. This was indeed observed with the first ever neutron diffraction experiment on MnO performed by Shull and Wollan in 1949 (37), which also happened to be the first direct evidence of antiferromagnetism as shown in **Figure 5a**. To explicitly derive how the magnetic symmetry influences the neutron scattering in MnO, one first notes that the peak with the lowest scattering angle at 80 K is at half the scattering angle of that at 293 K, indicating doubling of the crystallographic unit cell caused by the magnetic structure of the antiferromagnetic phase. For X-ray diffraction, the scattering intensity at a wavevector difference of $\Delta k$ is approximately given by $I_{XRD}(\Delta k) \propto \left| \sum_j f^j e^{-2\pi i r_j \cdot \Delta k} \right|^2$, where the summation is over the ions in the unit cell. Considering only the Mn atoms lattice sites but ignoring their spins, this then evaluates to $I_{XRD}(\Delta k) \propto \left| f^{Mn} e^{-2\pi i (0\ 0\ 0) \cdot \Delta k} + f^{Mn} e^{-2\pi i \left(\frac{1}{2}\frac{1}{2} 0\right) \cdot \Delta k} \right|^2$ which is non-zero when $(h, k, l)$ for the diffracted planes are either all-odd or all-even; however all-even indices are not experimentally observed above. For scattering of a neutron beam in the antiferromagnetic phase,

$$I_{ND}(\Delta k) \propto \left| f^{Mn\uparrow} e^{-2\pi i (0\ 0\ 0) \cdot \Delta k} + f^{Mn\uparrow} e^{-2\pi i \left(\frac{1}{2}\frac{1}{2} 0\right) \cdot \Delta k} + f^{Mn\downarrow} e^{-2\pi i \left(\frac{1}{4}\frac{1}{4} 0\right) \cdot \Delta k} + \right.$$

$$\left. f^{Mn\downarrow} e^{-2\pi i \left(\frac{1}{2}\ 0\ 0\right) \cdot \Delta k} \right|^2,$$ where $f^{Mn\uparrow} = -f^{Mn\downarrow}$ because of the interaction of neutrons with electron spins. The selection rules evaluate to $(h, k, l)$ being all-odd as observed experimentally in Figure 4(a). The Bragg diffraction of neutron beams at scattering angles that were believed to be forbidden based on the crystallographic space group symmetry ($Fm\overline{3}m$) of MnO and its associated X-ray

diffraction selection rules were actually allowed based on the magnetic space group ($C_{2c}2/m'$) of MnO.

### 3.2. Neumann's Principle Connects Magnetic Symmetry to Properties

In the context of magnetic properties such as magnetic susceptibility, magnetoelectricity, and piezomagnetism (2), it is this magnetic symmetry that is of interest when applying Neumann's principle stated in Section 2.1. To illustrate this, we consider the property of magnetoelectricity, $M_i = Q_{ij}E_j$, where $M_i$ represents the components of the magnetization, an axial time-odd vector, $E_j$ the components of the electric field, a polar time-even vector and $Q_{ij}$ the axial time-odd magnetoelectric second rank tensor. By Neumann's principle, the magnetoelectric tensor vanishes if $1'$ or $\overline{1}$ are symmetry elements of the point group of a crystal (2). Thus all 32 magnetic grey groups, and 11 colorless groups with explicit inversion symmetry lack magnetoelectric effect. The magnetoelectric effect was experimentally discovered for the first time (39) in chromium oxide, $Cr_2O_3$ with a magnetic point group of $Cr_2O_3$ is $\overline{3}'m'$. Neumann's principle implies then that $Q_{11} = Q_{22}$ and $Q_{33}$ are the non-zero elements (subscript 3 is along the 3-fold axis), both of which disappear above the magnetic transition temperature $T_N$ (**Figure 6a**).

However, care is needed in applying $1'$ (40). For example, consider Ohm's law, $J_i = \sigma_{ij}E_j$, where $i, j$ indicate Cartesian coordinates, $J$ is the current density in A/m$^2$, a time-odd vector, $E$ is the electric field in V/m, a time-even vector, and $\sigma$ is the electrical conductivity, a time-odd second rank tensor. Note that no magnetism is explicitly involved in this phenomenon, but time-reversal is involved. From Neumann's principle, one would conclude for a paramagnetic metal which explicitly contains $1'$ as a symmetry element, that $\sigma_{ij} = 0$, which is unphysical since paramagnetic metals are good conductors of electricity. The apparent contradiction is resolved by noting that the current flow leads to energy dissipation through heat, which breaks time reversal symmetry, $1'$. In the well-known Hall effect (41), before applying a magnetic field, the longitudinal conductivity, $\sigma_{ii}$, is non-zero (due to dissipation), but the transverse conductivity $\sigma_{ij}$, $i \neq j$ is zero, suggesting that while longitudinal dissipation is significant, the transverse dissipation in these cases might be ignored since no work is considered done when the field and current are in orthogonal directions. In such a case, time reversal $1'$ symmetry is preserved for the transverse conductivity but not for the longitudinal conductivity. However, when a magnetic field is turned on, $1'$ is broken and a transverse Hall conductivity is also observed as expected.

### 3.3. Time-Reversal in Classical versus Quantum Mechanics

As useful as it is to formulate time-reversal symmetry as an antisymmetry $1'$, this is in principle only valid under a purely classical treatment of spins as loops of current. Laws of classical mechanics are invariant under the transformation $t \rightarrow -t$, whereas in quantum mechanics, time reversal additionally involves a complex conjugation operation (42). Furthermore, for quantum mechanics with half-integer spin systems, time-reversal symmetry is no longer even a self-inverse transformation, which is one of the requirements for an operation to be an antisymmetry (see Section 2.2). Nonetheless, magnetic space groups are in many cases applied to truly quantum mechanical phenomena (43). Below, we clarify these subtle issues.

First, following Wigner, we think of time-reversal symmetry as a "reversal of the direction of motion"(44). **Figure 7** depicts the action of an operator $\mathcal{T}$ (it is identical to $1'$ used in crystallography) on position $x$ and momentum, $p$ as follows: $\mathcal{T}\big(x(t), p(t)\big) = \big(x(-t), -p(-t)\big)$. Laws of nature are invariant under time-reversal symmetry with some exceptions involving weak interactions in particle physics (45). For instance, $m\ddot{x} = F = -\nabla V(x)$, is invariant under time reversal, where $V(x)$ is the potential associated with the conservative force $F$, where both $x(t)$ and $x(-t)$ are solutions. Maxwell's equations are also invariant under time-reversal, if one considers the magnetic field $\boldsymbol{B}$ to arise from moving charges or currents, so that $\boldsymbol{B} \rightarrow -\boldsymbol{B}$ under $\mathcal{T}$. Note that the operator $\mathcal{T}$ is clearly an antisymmetry, since it is a self-inverse operator ($\mathcal{T}^2 = 1$) and commutes with rotations, rotoinversions and translations.

We now consider time-reversal in quantum mechanics. We first note that given the Schrödinger wave equation $i\hbar \frac{\partial \psi}{\partial t} = \left(-\frac{\hbar^2}{2m}\nabla^2 + V\right)\psi$, for every $\psi(x, t)$ that is a solution, there exists another solution $\psi^*(x, -t)$. In fact, it can be shown (see Supplementary Material for a derivation) using the same recipe that was followed in **Figure 7** for a classical time-reversal symmetric trajectory, that given an eigenstate $\psi(x, t)$, the time-reversed eigenstate is given by $\psi^*(x, -t)$. It appears then, that the time-reversal operator in quantum mechanics, labeled $\Theta$ here to make a distinction from the classical $\mathcal{T}$, is given by complex conjugation operation, $K$, along with time reversal, $\mathcal{T}$: $t \rightarrow -t$, i.e. $\Theta = \mathcal{T}K$. We can quickly check that this formulation is consistent with the classical limit where the momentum is reversed under time-reversal, by noting that momentum eigenstates $|p\rangle = e^{-ip.r}$ are transformed to $|-p\rangle = e^{ip.r}$ under time-reversal, so that the classical expectation value of momentum transforms as $p \rightarrow -p$. Note that $\Theta^2 = \mathcal{T}K\mathcal{T}K = 1$, since $\mathcal{T}$ and $K$ commute. This formulation of time reversal, namely, $\Theta$, can thus be considered as time reversal in quantum mechanics formulations without spin ½ particles.

Strikingly, when quantum mechanical time-reversal is applied to an electron, which is a spin-1/2 particle, it can be shown (see Supplementary Material for a derivation) that time reversal is equal to $\Theta_S = \sigma_y K$, where the subscript $S$ is for a spin ½ particle, and $\sigma_y$ is the 2×2 Pauli matrix (0 -$i$; $i$ 0). Noting that $K\sigma_y = -\sigma_y K$, it can now be shown that $\Theta_S^2 = \sigma_y K \sigma_y K = -1$. Therefore, if we apply time-reversal twice on say, a spin-up particle in a state $|\uparrow\rangle$, we obtain $\Theta_S^2|\uparrow\rangle = -|\uparrow\rangle$, and likewise with a spin-down particle, $\Theta_S^2|\downarrow\rangle = -|\downarrow\rangle$. This is a radical departure from what is expected from classical 'current loop' magnetic moments, but one with important consequences that are ubiquitous in quantum mechanics. The most prominent example is that of Kramers theorem (46–48) depicted in **Figure 8**, which states that for every Bloch state $|\psi_k\rangle$ in a crystal with spin $\downarrow$ and energy $\epsilon_k$, there exists an orthogonal eigenstate $\Theta_S|\psi\rangle$ with spin $\uparrow$ and the same energy, $\epsilon$, i.e. $\epsilon_{k,\uparrow}^m = \epsilon_{-k,\downarrow}^n$, where $k$ is the particle wavevector and the superscripts indicate two different energy bands $m$ and $n$. The band structure is thus symmetric about $k = 0$. Another consequence is that the eigenstates at time-reversal invariant momenta $k = 0$ and $k = \pm\frac{\pi}{a}$ are two-fold degenerate. This subset of $\boldsymbol{k}$-points are deemed time-reversal invariant momenta (TRIM), because applying the time-reversal operation $\Theta$ leaves each of them completely invariant to within a reciprocal lattice vector $\boldsymbol{G}$, i.e. $\boldsymbol{k} \overset{\Theta}{\rightarrow} -\boldsymbol{k} = \boldsymbol{k} + \boldsymbol{G}$. Kramers' degeneracy can be broken by breaking time-reversal symmetry, such as by applying an external magnetic field $B$.

Now we go back to our original question – It appears based on the result $\Theta_S^2 = -1$ for spin-1/2 particles, that $\Theta_S$ is no longer a self-inverse operation; can it then be still treated as an antisymmetry? In both quantum and classical mechanics, we can think of the time-reversal operator as flipping the sense of rotation of the 'tiny current loops' that defines the electron spin, but that said, the associated quantum phase that is inherited in the time-reversal operation can have physically meaningful results such as in the Aharonov-Bohm effect (49). It is easy to show (see Supplementary Material) that as long as one considers only expectation values of operators, rather than off diagonal matrix elements, the classical time-reversal operator $\mathcal{T}$ and quantum mechanical time-reversal operator $\Theta$ will give the same result. It can also be shown (50) that in the absence of spin-orbit coupling, Θ can be assumed to be simply $K$ even for a spin-1/2 particle. Clearly, $\Theta_S = K$ is a self-inverse operation since $K^2 = 1$. It is in these limits that the formulation of $1'$ as an antisymmetry, and the associated magnetic space groups are valid in the study of quantum phenomena as well. Further, instead of representation theory, we must use co-representation theory that is suitable for anti-unitary groups, a discussion of which is covered in detail by Cracknell and Bradley (50) and is beyond the scope of this review.

## 4. ANTISYMMETRY OF DISTORTIONS

A recent application of antisymmetry in material research comes in the form of *distortion reversal antisymmetry*, 1* that reverses the atomic trajectories that compose an arbitrary distortion. Consequently, a symmetry-based description of a wide range of phenomena is enabled including atomic diffusion, vibrations, phase transitions, interface dynamics, and ferroelectric and magnetic switching.

Distortion reversal traces its origin in the "rotation reversal" antisymmetry operation, $1^\Phi$, proposed by Litvin and Gopalan in 2011 (26) as illustrated in **Figure 9**. Gopalan and Litvin (26) identified the orange box in **Figure 9a** as an empty entry, and proposed to place $1\Phi$ there to reverse time-even axial vectors. It was introduced to reverse the sense of rotations of polyhedra composing a crystal, between clockwise ($-\Phi$) to counterclockwise ($+\Phi$) directions about the center of mass of each polyhedron, i.e. $1^\Phi: \Phi \rightarrow -\Phi$. This is shown schematically in **Figure 9b** and **9c** - polyhedra composed of black circles are rotated counterclockwise (CCW), while those of white circles are rotated clockwise, each by an angle $\Phi$. The antisymmetry operation was conceived as reversing the sign of $\Phi$, namely, $1\Phi: \Phi \rightarrow -\Phi$ corresponding to colors of black and white, respectively. Thus, this structure is identified to have a colorless point group symmetry of mm2, and a color symmetry of $4_z^{\Phi} m_x m_{xy}^{\Phi}$. Examples of applications included rotations of $O_6$ oxygen octahedra in complex oxides(51) and $SiO_4$ tetrahedra in silicate structures such as quartz (52). However, Vanleeuwen identified (27) the difficulty in uniquely identifying polyhedras in solids as illustrated in panels **d** and **e** of **Figure 9**. In addressing these shortcomings, arose the idea of distortion reversal symmetry, 1*, proposed by VanLeeuwen and Gopalan in 2015 (10). If the trajectories of atoms are deterministically tracked in the distortion process, and all polyhedral identification is dropped, then the distortion can simply be considered as a collection of trajectories of each atom, as shown in **Figure 9f**. Each trajectory can be parametrized by a dimensionless parameter, $-1 \le \lambda \le 1$, where -1 and +1 represent the two extrema of each trajectory as shown in black and white. This is described in detail in the next section. A similar concept called choreographic symmetry was proposed by Boyle, Khoo and Smith in 2016 (53), but it is not reviewed here separately.

## 4.1. Distortion Reversal Antisymmetry, 1*

In order to fix the arbitrary nature of choosing polyhedra and rotation angles, one should first avoid the need for any polyhedral identification in a solid and directly work with the atoms and their atomic trajectories involved in a distortion. A general *distortion*, whether a rigid body rotation, translation, scaling, or a general deformation involving changes in internal angles, can then be described by a collection of atomic trajectories as shown in panel **f** in **Figure 9**. By parameterizing the atomic trajectories traced out with a reaction coordinate ($\lambda$), an antisymmetry operation acting on $\lambda$ can be defined as distortion reversal antisymmetry: $1^*: \lambda \rightarrow -\lambda$; It reverses the individual atomic trajectories which make up the pathway of the distortion. Since $\lambda$ is a general reaction coordinate, many different aspects of a system can be used to parameterize with it.

**Figure 10** illustrates a simple example of a ball rolling down a hill where its height is used to parameterize the process linearly with λ. Here the top of the hill (initial state) is at λ = -1 and the bottom of the hill is at λ = +1. In this example, it is important to note that λ evolves *quadratically* with time (*t*) as the ball accelerates down the hill due to gravity. Both 1′ and 1* for this process act in different spaces, respectively, reversing the time (*t*) and "time-like" (λ) coordinates independently. Conventionally, λ = -1 is taken as the initial state, and λ = +1 as the final state. Under the action of 1*, the initial and final states are reversed.

## 4.2. Distortion Symmetry Groups Uniquely Tag a Distortion Path

The antisymmetries 1* and 1′ can be combined with the 32 crystallographic point groups and 230 space groups to form 624 DAPG and 17,803 DASG (see Section 2.2). Of these, the groups that do not explicitly contain 1* (non-grey groups) were named by VanLeeuwen and Gopalan (10) as *distortion symmetry groups*; this was in analogy with the magnetic groups that do not contain 1′ explicitly. These consist of a set of elements that leave a whole path invariant when applied to its atomic trajectories. Consequently, it is thus the collective symmetry of an entire path instead of the symmetry of the individual frames (or images) within the path. Distortion symmetry groups are useful in finding minimum energy pathways (MEPs) between an initial and a final state of a material system. This first requires determining the distortion symmetry group of an initial path.

To tag a path with a unique distortion group, *G*, we first determine the group of "unstarred" symmetry operations (ones that do not have any association with the distortion reversal operation, 1*) that leave the whole path invariant. This group (*H*) can be obtained through an intersection of the conventional crystallographic symmetry groups $S(\lambda)$ of each frame λ in a path, i.e $H = \bigcap_{-1 \le \lambda \le 1} S(\lambda)$, that ranges by convention as stated earlier from λ = -1 and λ = +1. From here on, a set of elements in $S(0)$ are identified (call them *A*), which map the system at λ to − λ. When combined with 1*, i.e. 1*$A$, the new elements generated will be symmetries of the path. The elements of the overall distortion group $G$ of the entire path can then be obtained as, $G = H \cup 1^*A\,H$. **Figure 11** illustrates obtaining the distortion group for the path of an oxygen atom diffusing across a graphene surface. Additionally, since one has the control in choosing the initial path in finding a MEP, it can be highly symmetrized by choice in order to take full advantage of distortion symmetry and group theory; one can systematically lower the symmetry with group theory dictated perturbations and find many intermediate paths starting from a single path.

### 4.3.  Finding Minimum Energy Pathways (MEPs) using Distortion Symmetry Groups

Distortion groups have recently been demonstrated to be a powerful tool for finding minimum energy pathways (MEPs), identified as the lowest energy path for the distortion (rearrangement) of a group of atoms from an initial stable state ($\lambda = -1$) to a final stable state ($\lambda = +1$). The reaction coordinate here is thus, $\lambda$, whether it is temperature, pressure, stress, electric or magnetic fields, or any other intensive parameter in response to which the system changes state. The potential barrier maximum then corresponds to a saddle point energy which determines the transition rate within the harmonic transition state theory (54).

There are infinitely many paths from an initial to a final state in the phase space of intensive parameters mentioned above. Determining the MEP from among them is thus a central problem in all of materials science, chemistry, physics and biology.  Of the many methods (55, 56) that are used to find *an* MEP (instead of *the* MEP, since Nudged Elastic Band method (NEB) cannot ensure a global minimum, but only a local minimum) starting from an initial guess path, the NEB is one of the most common(57–59). The essential idea is depicted in **Figure 12**. The NEB is a chain-of-states method, where an initial guess path between the initial and final states is defined by a series of intermediate frames (called images) which are artificially connected by springs of some stiffness (not shown) to keep them uniformly apart along the path. This in turn allows for the parametrization of the images with a uniformly distributed λ. In NEB, the forces on these images are then put through a projection scheme such that the (artificial) spring forces perpendicular to the path ($\boldsymbol{F}_{\perp}^{spring}$) (shown in the inset) and the component of the true force (equal to the gradient of the energy potential) parallel to the path (i.e. $-\boldsymbol{\nabla}V_{\parallel}$) are zero; this process is called "nudging". Thus a path, the surviving force on any image $i$ is given by $\boldsymbol{F}_i^{NEB} = -\boldsymbol{\nabla}V(\lambda_i)_{\perp} + \boldsymbol{F}_{i,\parallel}^{spring}$. From here, an optimization scheme can then be employed to move the images along $\boldsymbol{F}_i^{NEB}$ and try and converge to an MEP. The process of finding the true MEP is however stochastic, relying on guessing as many initial starting paths as possible and comparing their NEB optimized MEPs to find *the* most likely MEP. No symmetry principles are used in NEB or any other numerical method to find MEPs. Lack of numerical convergence in running NEBs also results in one never being sure if a final MEP is correct or even properly converged (see examples in the supplementary section of Reference (10)).

Distortion symmetry allows for a tagging of each possible distortion path with a unique distortion group (Section 4.2). Further, by picking an initial path, and determining its distortion group, group theory tells us the number of irreducible representations (irreps) in that group. Each 1-D irrep represents a unique way to perturb the initial path, and each irrep is orthogonal to the others. Irreps with $n$-dimensions ($nD$) will produce $n$ perturbation vectors as discussed further in the next section. In this manner, *there are only a finite number of distinct ways to perturb an initial path* as determined by the distortion group and its group theory representation. Each irrep can then be used to perturb the initial path and lower its distortion symmetry systematically (28, 29). Once a MEP is found, its distortion group will determine the unique number of ways to perturb that path further. Once a true (local) MEP is found by the NEB, any further perturbations based on its irreps will not lower the symmetry of the path any further. This is because, the *distortion symmetry of a path can only be unchanged or raised by the NEB calculation, but never lowered* (10). In other words, if $S$ is a symmetry of a path $P$, i.e. if $SP=P$, then S [NEB($P$)] = NEB($P$).

### 4.4. DiSPy: Distortion Symmetry Method Implemented in Python

The above procedure has been deemed the *distortion symmetry method* (DSM)*,* and has recently been implemented into a Python package (DiSPy) by Munro and coworkers (29) and applied to the study of vibrations, diffusion(10) and ferroelectric switching (28). In doing so, it was demonstrated how the tool can enable previously overlooked switching pathways to be discovered, as well allow for the exploration of the potential energy landscape around an initially chosen highly symmetric path.

The DiSPy package works by using projection operators to generate symmetry-adapted perturbations from what would be an arbitrary perturbation to an initial path. This is effective because the goal of any path perturbation is to invoke unstable modes of an initial path that may exist to lower its symmetry. Since these unstable displacive modes necessarily transform as irreps of the path's distortion symmetry group, access to them is granted through the use of these operators. More specifically, individual displacive modes can be generated with projection operators which form a basis for any stable or unstable displacive mode that transforms as that same irrep. Perturbing with these can then be used to push a path along its instabilities without having to directly calculate the modes associated with them. This is advantageous, as this type of calculation for many systems would be computationally expensive to perform. More specific details on the theoretical background of the method can be found in the resource by Munro and coworkers (29). The goal of the DiSPy package is to make it easy to generate symmetry-adapted perturbations as described above. The package is written to use the input and output capabilities included in the pymatgen package (60), thus accommodating a large variety of formats for many electronic structure and molecular dynamics packages.

## 4.5. Examples of MEPs using Nudged Elastic Band Method + Distortion Symmetry

The above approach can be applied to the diffusion example shown in **Figure 11**. If the NEB algorithm is run using the linearly interpolated path (**Figure 11a**), a high energy pathway is obtained (**Figure 11d**). Using DiSPy, this relaxed path can then be perturbed with symmetry-adapted perturbations constructed using the irrep matrices shown in the character table of $m^*m2^*$ (**Figure 11c**). By using the perturbation associated with the $\Gamma_2$ irrep, a lower energy path with a distortion symmetry of $m^*$ is found if the NEB algorithm is run again. This lower energy path instead consists of the oxygen diffusing along the edge of the carbon ring instead of across it. It should be noted that perturbations generated with the other two non-trivial irreps simply cause NEB to return back to the high energy linearly interpolated path. In other words, they are 'stable' perturbations.

The DSM (using the DiSPy code) was applied to the study of ferroelectric switching in $Ca_3Ti_2O_7$, an improper ferroelectric, $PbTiO_3$ and $LiNbO_3$, a proper ferroelectric, and $BiFeO_3$, a multiferroic (28, 29). For $Ca_3Ti_2O_7$ and $BiFeO_3$ in particular, low energy switching pathways were discovered beyond those reported in the literature using conventional approaches with NEB calculations (61, 62). In $Ca_3Ti_2O_7$, six new four-step paths were found which pass through previously known low energy orthorhombic and monoclinic phases of the material (**Figure 13**) (28). In $BiFeO_3$, switching involving both the polarization and magnetization was explored, defining two different kinds of paths – those which switch only the polarization but not the magnetization, and others which switch both. As a result, it was shown how the polarization of the material could be reversed without reversing the net magnetization, at a similar energetic cost to the process when both are reversed (28). This informs studies where deterministic switching of the net magnetization through control of the polarization is sought (63). Interestingly, the newly discovered low energy path was also found to have a unique intermediate metastable structure where the polarization and magnetization in the BiFeO$_3$ were aligned parallel or antiparallel to each other; normally they are perpendicular to each other in the ground state (64).

If the initial guess path for NEB is selected to be one with a very high degree of symmetry, many different symmetry-adapted perturbations can be calculated and used to explore a large number of potential pathways. An example of this is given by the study on pathways for domain wall motion in PbTiO$_3$ (28). The DSM was used to obtain many different pathways between the two end states, and the polarization at the domain wall was calculated as a function of the reaction coordinate for each of the paths. Unlike the classical expectation of Ising ferroelectric walls, finite Bloch and Ne*é*l components (65) of polarization arose in many of the newly obtained paths.

## 5. ANTISYMMETRY-BASED CLASSIFICATION OF PHYSICAL QUANTITIES

### 5.1. Current Classification of Physical Quantities in 3D

*Can we classify all (non-relativistic) physical quantities in arbitrary dimensions?* Physical quantities are typically represented in terms of scalars, vectors and tensors of various types, collectively called "vectorlike" quantities by Hlinka (66). **Figure 14** depicts 8 types of *"vector-like"* objects in 3D as classified in Ref. (66): (1) *Neutral* (scalars) of two types: time-even (such as charge $q$ or length $r$), or time-odd (such as time, $t$), (2) *Polar* (vectors) of two types: time-even (such as position vector $\boldsymbol{r}$) or a time-odd (such as velocity vector, $\boldsymbol{v}$), (3) *Axial* (vectors) of two types: time-even (such as a polarization loop, $\boldsymbol{r} \times \boldsymbol{P}$) or time-odd (such as current density loop, $\boldsymbol{r} \times \boldsymbol{J}$), where $\boldsymbol{r}$ is the radius vector of the loop and $\boldsymbol{P}$ and $\boldsymbol{J}$ are tangential polarization and current density vectors, respectively around the perimeter of the loop, and (4) *Chiral* "vectors" of two types: time-even (such as helical winding, $(\boldsymbol{r} \times \boldsymbol{T}) \cdot \boldsymbol{n}$) or a time-odd (such as current moving through a solenoid, $(\boldsymbol{r} \times \boldsymbol{J}) \cdot \boldsymbol{n}$), where $\boldsymbol{r}$, $\boldsymbol{T}$, $\boldsymbol{n}$ are respectively, the radial, tangential and axis vectors of the helix as shown in **Figure 14**. In Hlinka notation, the first four objects from left to right would respectively be named N, P, G, C for time-even, and L, T, M, F for time-odd objects. Note that axial vectors (also called pseudovectors in 3D) can be written as cross-products between two polar vectors. The chiral "vectors" are really pseudoscalars in 3D and involve a dot product between an axial and a polar vector. **Figure 14** also indicates the transformation of these objects under $\bar{1}$, $1'$, and a mirror plane $m_{\parallel}$ in 3D space defined in Ref (66). Here time-even refers to invariance under time reversal, $1'$, and time-odd refers to reversal of the object or quantity under $1'$. Similarly, $\bar{1}$-even is termed *centric* and $\bar{1}$-odd is termed *acentric*.

## 5.2. Transitioning from 3D to $n$D using the language of Multivectors

Note that the above quantities, e.g., $q$, $\boldsymbol{r}$, $\times \boldsymbol{P}$, $(\boldsymbol{r} \times \boldsymbol{T}) \cdot \boldsymbol{n}$, depicted in **Figure 14** are composed of zero, one, two and three vectors, respectively. One could continue to define more physical quantities composed of an even larger number of vectors, and thus an infinite sequence of quantities could be composed as scalar S (grade 0 composed of no vectors), vector V (grade 1 composed of one vector), bivector B (grade 2, composed of two vectors), trivector T (grade 3), quadvector Q (grade 4), pentavector P (grade 5), and so on for a general *blade* (of *grade g* composed of $g$ vectors), as they are called in Clifford Algebra (CA) (see Supplementary Material for a brief introduction to CA). In addition, CA allows the addition of such blades, for example, M=S+V+B+T+Q+P etc. which is an arbitrary *multivector* living in a $2^5$-dimensional CA space aising from a 5-dimensional vector space. For example, in CA, the electromagnetic field F is a multivector defined as F=$\boldsymbol{E}$+$c$B, where $\boldsymbol{E}$ is the electric field vector and B is the magnetic field *bivector*. Similarly, the current density J=$(\rho/\varepsilon_o) - c\mu_o\boldsymbol{J}$, is a multivector (charge density, $\rho$, $c$ is speed of light in vacuum, permittivity $\varepsilon_o$ and permeability $\mu_o$ of free space).   CA allows one to write all four of Maxwell's equations in free space succinctly as one single equation, ($\boldsymbol{\nabla} + [\frac{1}{c}]\,\partial/\partial t)$F =J in the Newtonian space plus scalar time, *t*, a process called "encoding" that reveals deeper interconnections between diverse laws (12, 67).

This may seem unusual at first, because one is typically more conversant with physical properties being conventionally represented in the language of *tensors* of different *ranks*, such as scalars (rank 1), vectors (rank 1), and tensors of ranks 2, 3, 4…etc. Further, adding scalars and vectors and tensors of higher ranks is unusual.  One can however show that every tensor can be written as a multivector, and hence the language of multivectors is an alternate representation of tensors, but with the advantage of being "coordinate-free"  (68, 69).  As a simple example, consider a second rank tensor property, $T_{ij}$ in 2-dimensions spanned by unit vectors $\hat{\boldsymbol{x}}$ and $\hat{\boldsymbol{y}}$, which thus has 4 independent terms, $T_{11}$, $T_{12}$, $T_{21}$ and $T_{22}$.  One can write this tensor as a multivector $M$, where $2M = (T_{11} + T_{22}) + (T_{12} + T_{21})\hat{\boldsymbol{x}} + (T_{11} - T_{22})\hat{\boldsymbol{y}} + (T_{12} - T_{21})\hat{\boldsymbol{x}}\hat{\boldsymbol{y}}$.  While the first term on the right is a scalar, and the next two are vectors, the last term appears to be a new type of axis represented by the unit vector $\hat{\boldsymbol{x}}\hat{\boldsymbol{y}}$.  Indeed, it is in CA: $\hat{\boldsymbol{x}}\hat{\boldsymbol{y}}$ is a *unit bivector*, which is formed from the *geometric product* between two unit-vectors $\hat{\boldsymbol{x}}$ and $\hat{\boldsymbol{y}}$.  (See Supplementary Material).  In particular,  $\hat{\boldsymbol{x}}\hat{\boldsymbol{y}}$ represents a unit area in the $\hat{\boldsymbol{x}}$-$\hat{\boldsymbol{y}}$ plane. It has a clockwise circulation of vectors $\hat{\boldsymbol{x}}$ and $\hat{\boldsymbol{y}}$ around its perimeter (see **Figure 15**). Similarly, $\hat{\boldsymbol{y}}\hat{\boldsymbol{x}}$ represents a unit area with a counterclockwise circulation of vectors around its perimeter, such that $\hat{\boldsymbol{x}}\hat{\boldsymbol{y}} = -\hat{\boldsymbol{y}}\hat{\boldsymbol{x}}$, which is the condition for orthogonality between two vectors in CA.  If the reader is wondering how one is to "multiply" two vectors, $\hat{\boldsymbol{x}}$ and $\hat{\boldsymbol{y}}$ in this fashion, one can for example, define $\hat{\boldsymbol{x}} = (0\ 1;\ 1\ 0)$, and $\hat{\boldsymbol{y}} = (1\ 0;\ 0 - 1)$, i.e. as two 2×2 Pauli matrices. Then, $\hat{\boldsymbol{x}}\hat{\boldsymbol{y}} = (0 - 1;\ 1\ 0) = -\hat{\boldsymbol{y}}\hat{\boldsymbol{x}}$. Also note that $\widehat{\boldsymbol{x}}\widehat{\boldsymbol{x}} = \widehat{\boldsymbol{y}}\widehat{\boldsymbol{y}} = (1\ 0; 0\ 1) = I$, a unit matrix. Starting from a two-dimensional ($n$=2) vector space spanned by $\widehat{\boldsymbol{x}}$, $\widehat{\boldsymbol{y}}$, one is thus led to a $2^n$= 4-dimensional CA space of $I$, $\widehat{\boldsymbol{x}}$, $\widehat{\boldsymbol{y}}$, and $\widehat{\boldsymbol{x}}\widehat{\boldsymbol{y}}$, composed of a scalar axis spanning the real number subspace, two vector axes spanning a vector subspace, and a bivector axis defining its own subspace. This can be generalized to any dimension (see Supplementary Material). In this manner, all of the physical properties conventionally expressed as tensors can be rewritten as multivectors in CA. By doing so, Gopalan (11) recently showed that there are only 41 types of multivectors representing physical quantities in arbitrary dimensions.

### 5.3. Dropping Axial and Chiral traits in generalizing to $n$-dimensions

To classify multivectors in general, however, the traits of *axiality* and *chirality*, and as a consequence the mirror operation, $m_\parallel$ in **Figure 14** need to be dropped. The idea of axial vectors is not generalizable beyond 3D (with the curious exception of 7D, as an aside) (70). For example, in a 2D ambient space, the 2D loop in **Figure 14** has no normal at all, while in a 4D ambience, it has two normals! Thus *axial vectors* have to be dropped in favor of *bivectors*. Chirality of an object in dimension $n$ is tested conventionally by creating its mirror image using a mirror (a hyperplane) of dimension $n$-1, and looking for congruent overlap between the original and the mirror images; if they can be congruently overlapped, the $n$D object is *achiral* in the $n$D space, and otherwise, the $n$D object is *chiral* in $n$D. Now referring to **Figure 16**, in an $n$-dimensional space, only an $n$-dimensional object can be chiral. However, the same $n$D object will become achiral in a space of dimensionality ($n$+1) or higher. We thus conclude that the trait of chirality is also not unique to an object without reference to the dimensionality of its ambient space. Next, we describe *wedge reversion*, $1^\dagger$, proposed recently by Gopalan (11) that resolves both these issues.

## 5.4. Wedge Reversion, $1^\dagger$, as the Missing Antisymmetry

Wedge reversion is not a new operation; it is simply called *reverse*, or *reversion* in Clifford Algebra (68, 69). What is new in the recent work by Gopalan (11) is that it is being formally given the status of an antisymmetry, $1^\dagger$. To define $1^\dagger$, first, we need to define a wedge product between two vectors $\boldsymbol{A}$ and $\boldsymbol{B}$ as $\boldsymbol{A} \wedge \boldsymbol{B}$ (see Supplementary Material for a formal definition of wedge product). For example, if $\boldsymbol{A} = 5\hat{\boldsymbol{x}} + 2\hat{\boldsymbol{y}}$ and $\boldsymbol{B} = 2\hat{\boldsymbol{x}} - 3\hat{\boldsymbol{y}}$, then $\boldsymbol{AB} = 4I - 19\hat{\boldsymbol{x}}\hat{\boldsymbol{y}}$, where we have made use of the orthonormality conditions (see Section 5.2 and Supplementary Material) that $\hat{\boldsymbol{x}}\hat{\boldsymbol{x}} = \hat{\boldsymbol{y}}\hat{\boldsymbol{y}} = I$, and $\hat{\boldsymbol{x}}\hat{\boldsymbol{y}} = -\hat{\boldsymbol{y}}\hat{\boldsymbol{x}}$. In this example, $\boldsymbol{A} \cdot \boldsymbol{B} = 4I$ (a scalar) and $\boldsymbol{A} \wedge \boldsymbol{B} = -19\hat{\boldsymbol{x}}\hat{\boldsymbol{y}}$ (a bivector). Note that $\boldsymbol{A} \wedge \boldsymbol{B}$ (a bivector) is different from the conventional cross-product, $\boldsymbol{A} \times \boldsymbol{B} = -19\hat{\boldsymbol{z}}$ (a vector); the former lives in the bivector subspace while the latter lives in the vector subspace. Also note that $\hat{\boldsymbol{x}}\hat{\boldsymbol{y}} = \hat{\boldsymbol{x}} \wedge \hat{\boldsymbol{y}}$ etc.

Geometrically speaking, the magnitude of $\boldsymbol{A} \wedge \boldsymbol{B}$ is the area of the parallelogram formed by vectors $\boldsymbol{A}$ and $\boldsymbol{B}$; the two vectors define a sense of "circulation" around the edges of this area. This sense of circulation is reversed in $\boldsymbol{B} \wedge \boldsymbol{A}$. Note that there is no reference to the dimensionality of the ambience of this object, and only to the dimensionality of the object itself; hence wedge product is generalizable. Similarly, the wedge product between three linearly independent vectors $\boldsymbol{A} \wedge \boldsymbol{B} \wedge \boldsymbol{C}$ gives the 3D volume of the parallelepiped enclosed by the three vectors. To generalize, the wedge product between $n$ linearly independent vectors gives the hypervolume enclosed by those vectors in $n$D.

Wedge reversion, $1^\dagger$, simply reverses the order of vectors in a wedge product as depicted in **Figure 17**. In particular, consider the 3D Euclidean space spanned by the orthonormal basis vectors, $\hat{\boldsymbol{x}}$, $\hat{\boldsymbol{y}}$, and $\hat{\boldsymbol{z}}$. Then, $1^\dagger(\hat{\boldsymbol{x}}) = \hat{\boldsymbol{x}}$, $\quad 1^\dagger(\hat{\boldsymbol{x}} \wedge \hat{\boldsymbol{y}}) = \hat{\boldsymbol{y}} \wedge \hat{\boldsymbol{x}} = -\hat{\boldsymbol{x}} \wedge \hat{\boldsymbol{y}}$, $\quad 1^\dagger(\hat{\boldsymbol{x}} \wedge \hat{\boldsymbol{y}} \wedge \hat{\boldsymbol{z}}) = \hat{\boldsymbol{z}} \wedge \hat{\boldsymbol{y}} \wedge$ $\hat{x} = -\hat{x} \wedge \hat{y} \wedge \hat{z}$.  These relations follow from the orthonormality conditions given in the Supplementary Material.  Thus, wedge reversion $1^\dagger$ leaves the scalars and the vectors invariant, but reverses the bivector, $\hat{x} \wedge \hat{y}$, and the trivector, $\hat{x} \wedge \hat{y} \wedge \hat{z}$.  Generally, $1^\dagger$ will reverse the sign of multivectors of grades $(4g+2)$ and $(4g+3)$, while leaving the multivectors of grades $4g$ and $(4g+1)$ invariant, where $g$=0, 1, 2, 3… is a whole number.  The fact that $1^\dagger$ will reverse the sign of some multivectors while not reversing that of others, led Gopalan (11) to use the term wedge *reversion*, rather than wedge *reversal*.

A note of caution regarding the action of $1^\dagger$ on axial vectors versus bivectors: the wedge reversion, $1^\dagger$, will reverse a bivector, $\boldsymbol{A} \wedge \boldsymbol{B}$, but not the vector, $\boldsymbol{A} \times \boldsymbol{B}$, which is consistent with the discussion above that grade 0 multivectors are invariant under $1^\dagger$, and emphasizes that there is only one type of vector, not several.  A similar word of caution exists between a trivector and a pseudoscalar in 3D: the action of $1^\dagger$ on a pseudoscalar, $(\boldsymbol{A} \times \boldsymbol{B}) \cdot \boldsymbol{C}$ is no different from its action on a conventional scalar, $s$, i.e. they are both invariant under the action of $1^\dagger$.

## 5.5. Classification of Multivectors based on the Actions of $\bar{1}, 1'$ and $1^\dagger$

Using the traits of *centric* ($\bar{1}$-even) versus acentric ($\bar{1}$-odd), and *circulant* ($1^\dagger$-odd) versus *acirculant* ($1^\dagger$- even) blades, Gopalan(11) classifies multivectors as shown in **Figure 18**. (The phrase "circulant" avoids the term "axial" for reasons mentioned in Section 5.2.) As intuitive examples, the first four types of objects in **Figure 18** are *centric-acirculant* (scalar), *acentric-acirculant* (vector), *centric-circulant* (bivector), and *acentric-circulant* (trivector). Each one can be time-even or time-odd, leading to 8 types of multivectors.

**Table 1** lists the 8 principle types of multivectors that are either invariant or reversed under the action of $\bar{1}, 1'$ and $1^\dagger$.  Physical quantities and properties can be expressed as multivectors, hence this classification is broadly applicable to classifying all properties. The Table also lists some examples of such physical properties. If the action of these antisymmetries on a multivector can also be mixed (i.e. neither even, nor odd), then there will be a total of 41 types of multivectors as listed recently by Gopalan (11) (see Table 1 in that reference). The multivectors presented in **Table 1** here however are termed "principal" because all other multivector types arise from various sums of these principal multivectors. Since tensors can be expressed as multivectors, this classification suggests that there are 8 principal and 41 overall types of tensors or tensor components that can be expressed as multivectors.

## 6. SUMMARY AND OUTLOOK

### 6.1. Why are there so few Antisymmetries?

It is remarkable that an abstract but simple idea such as an antisymmetry that switches between two states of a trait can have such practical applications in materials research and in physical sciences at large. Antisymmetry operations reviewed here such as between forward versus backwards time or spins (time reversal, $1'$ or $\mathcal{T}$ classically, and $\Theta$ and $\Theta_S$ in quantum mechanics), positions in space between $\boldsymbol{r}$ to $\boldsymbol{-r}$ (inversion $\bar{1}$, in the context of proper rotations), forward versus reverse motion of atoms in a distortion path (distortion reversal, $1^*$), and reversing the circulation of $n$ linearly independent vectors that define a hypervolume in $n$ dimensions (wedge reversion, $1^\dagger$) can lead to powerful ways of describing the structure and symmetry of materials, physical quantities and distortion paths, and to a wide variety of very practical applications in crystallography, describing physical laws, polar and magnetic materials, quantum mechanics, general distortions such as diffusion, vibrations, phase transitions, polar and magnetic domain switching, finding minimum energy pathways, mapping energy surfaces, and in the classification of multivectors.

Given the simplicity and usefulness of these antisymmetries, it is also equally surprising that since 1892 when the crystallographic groups were first listed, only a handful of antisymmetries reviewed here have been currently used in materials research. It is the belief of the authors that there must be many more such useful antisymmetries that are yet to be discovered if more researchers were to focus on the topic. By the same token, the topic of permutation symmetry groups formed from more than two colors (18, 71, 72) (see **Figure 2b** for example) could be more widely exploited in materials research, another area of opportunity for materials researchers.

## 6.2. Symmetry and Topology

An area of great current and growing interest is that of topological phenomena in materials research with many excellent reviews on the topic (48, 73–75). We briefly note the relationship between symmetry and topology which are two closely related concepts, and in particular, the role of antisymmetries $1'$, $\bar{1}$, and $1^*$ in determining topological phenomena.

Symmetry operations, namely, rotations, rotoinversions, translations, as well as antisymmetry operations are "distance-preserving', i.e. they preserve the norm (length) of a vector in the object on which they are operating. Thus, the internal angles and scaling are preserved under the action of symmetry operations; the object is self-congruent before and after the operation is performed. In contrast, a topological operation allows for a continuous adiabatic deformation of the object, as long as no cuts and stitches are made in the object. The object need *not* be self-congruent before and after the topological operation is performed.

The concept of symmetry plays a prominent role in topologically non-trivial states of matter. Here, symmetries of the material preserve the topological characteristics which result in distinct "symmetry-protected" topological phases. In other words, the Hamiltonian of a material that exhibits topological character cannot be adiabatically deformed to one that does not without breaking the protecting symmetry (73). One of the most common types of these phases studied are those protected by the antisymmetry of time-reversal. More specifically, topological insulators with this property have become a very popular topic of study in recent years (48, 73–75).

The protection of topological states in insulators through time-reversal symmetry ($\Theta_S$) is a result of the Kramer degeneracies that arise for Bloch states in the bandstructure of the material (See Section 3.3 and **Figure 8**). Kramer's degeneracy ensures that a set of wavevectors ($\boldsymbol{k}$-points) are deemed time-reversal invariant momenta (TRIM), as applying the time-reversal operation $\Theta_S$ leaves each of them completely invariant to within a reciprocal lattice vector $\boldsymbol{G}$. In two-dimensions four unique TRIM points exist, while in three-dimensions there are eight. It can be shown that consideration of these TRIMS is enough to calculate the indices of the $Z_2$ ($v_0; v_1, v_2, v_3$) invariant a quantity that characterizes the topological properties (47, 76, 77). A non-zero $v_0$ index indicates a "strong" topological insulator with surface states that are robust to disorder. Non-zero values of the other indices indicate "weak" topological characteristics where the same states vanish in the presence local changes to the structure that break translational symmetry. In the presence of inversion symmetry, $\bar{1}$ , the so-called Fu-Kane parity criteria (76) is widely used to easily characterize if a given centrosymmetric material is a topological insulator.

Finally, we note a clean result connecting distortion reversal antisymmetry, 1* and the Berry phase (46), an important topological parameter in phenomena such as Aharonov-Bohm effect (49) and the modern theory of polarization (78) in crystals. VanLeeuwen and Gopalan (10) showed that if a topological distortion path is distortion reversal symmetry invariant, i.e. if the path has explicit 1* symmetry, then the Berry phase will be identically zero. This result illustrates the power of an antisymmetry to obviate the need for further computation or measurements. Distortion groups are currently not employed in the field of topology, but we believe that they could play an important role in uniquely tagging each topological distortion path, and bringing the power of group theory to bear on the problems of classifying and discovering new topological phenomena.

## DISCLOSURE STATEMENT

The authors have no competing interests to report.

## ACKNOWLEDGMENTS

V.G. would like to acknowledge wonderful discussions with Daniel B. Litvin, Brian VanLeeuwen, Chaoxing Liu, Hirofumi Akamatsu, Mantao Huang, Vincent S. Liu, Yin Shi, Long-Qing Chen, Craig J. Fennie and Mikael C. Rechtsman. The concepts of distortion reversal and wedge reversion would not have come this far without these valuable interactions. V.G. would like to thank the late Robert E. Newnham for his gift of hand-crafted wooden octahedra connected with ball and socket joints which triggered the original rotation reversal symmetry idea. In writing this article, V.G., I. D., and H.P acknowledge support from the National Science Foundation grant number DMR-1807768 and the NSF-MRSEC Center for Nanoscale Science at Penn State through grant number DMR-1420620.

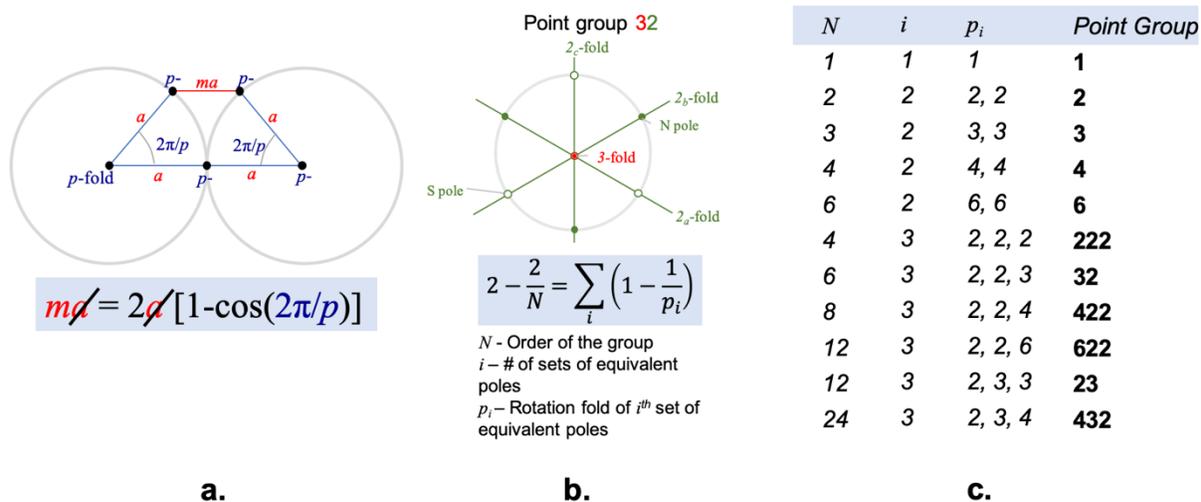

**a.**

$$m\cancel{a} = 2\cancel{a}/[1\text{-}\cos(2\pi/p)]$$

**b.**

Point group **32**

$$2 - \frac{2}{N} = \sum_{i}\left(1 - \frac{1}{p_i}\right)$$

*N* - Order of the group
*i* – # of sets of equivalent poles
$p_i$ – Rotation fold of $i^{th}$ set of equivalent poles

**c.**

| N | i | $p_i$ | Point Group |
|---|---|-------|-------------|
| 1 | 1 | 1 | **1** |
| 2 | 2 | 2, 2 | **2** |
| 3 | 2 | 3, 3 | **3** |
| 4 | 2 | 4, 4 | **4** |
| 6 | 2 | 6, 6 | **6** |
| 4 | 3 | 2, 2, 2 | **222** |
| 6 | 3 | 2, 2, 3 | **32** |
| 8 | 3 | 2, 2, 4 | **422** |
| 12 | 3 | 2, 2, 6 | **622** |
| 12 | 3 | 2, 3, 3 | **23** |
| 24 | 3 | 2, 3, 4 | **432** |

## Figure 1

Derivation of rotational crystallographic point groups. Panel (a) depicts the crystallographic restriction theorem, and shows a *p-fold* axis that periodically repeats every unit distance of $a$, and requiring the condition stated below, where $m$ is a whole number. Panel (b) poses the question as to how many *p-fold* axes can pass through a single pivot point at the center of an imaginary sphere (grey circle) and still form a group describing the symmetry of a crystal placed at the pivot point. The example of the point group 32 is the set of elements $\{1, 3, 3^{-1}, 2_a, 2_b, 2_c\}$; here $N = 6$ is the order of the group. Each axis has a north (N, filled) and a south (S, open) pole, where the rotation axes and the sphere intersect. Equivalent poles can be moved into congruence with each other using elements of the group. From the Table in (c), for the point group 32, $i = 3$ corresponding to the three sets of poles (∘,∘,∘), (·, ·, ·), (∘,·) as shown, with $p_i = 2$, 2, and 3, respectively, for the folds of the axes corresponding to each set of equivalent poles.

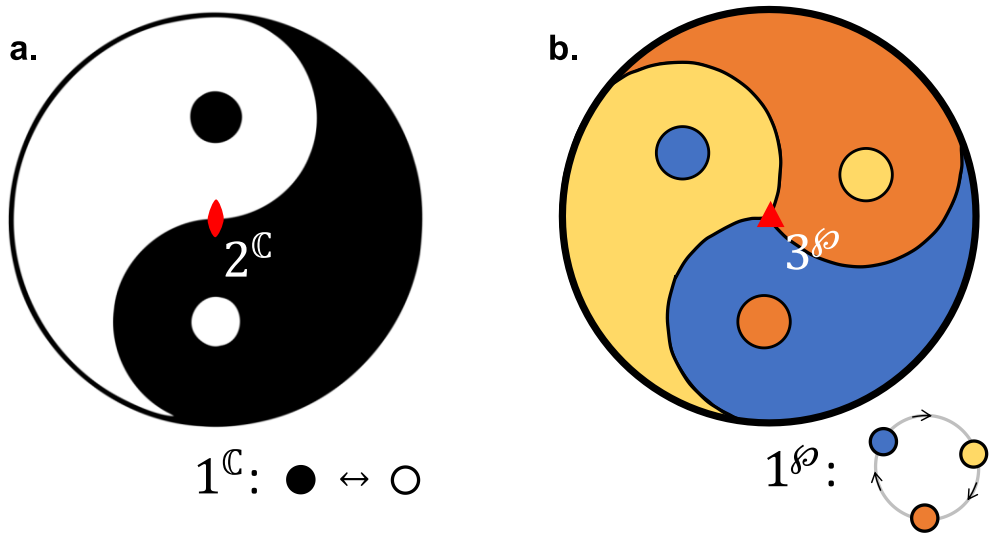

Color symmetry is illustrated with 2-color and 3-color objects. (*a*) If $1^{\mathbb{C}}$ is defined as an operation that exchanges black and white colors, then the Yin-Yang symbol in panel (a) is invariant under the symmetry operation $2^{\mathbb{C}}$, which is a 2-fold rotation followed by color swap. (*b*) If $1^{\wp}$ is defined as an operation that cyclically permutes orange-blue-yellow colors, then the 3-color symbol in panel (b) is invariant under the symmetry operation $3^{\wp}$, which is a 3-fold rotation followed by a three-color permutation.

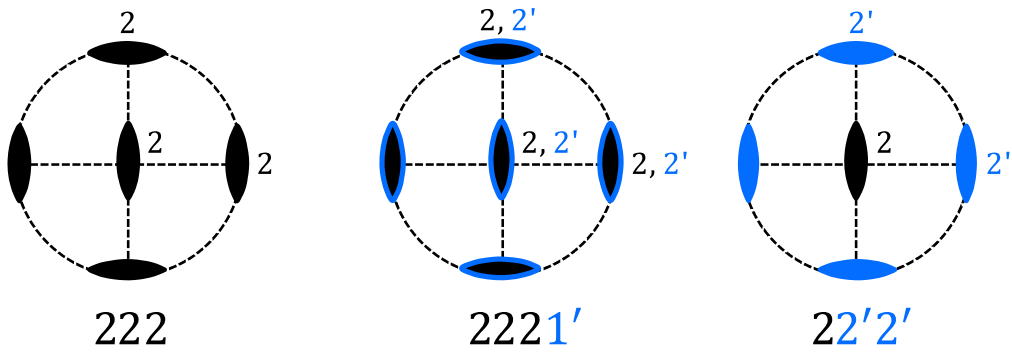

Figure 3

Stereographic projection of single antisymmetry point groups (SAPG). The groups are denoted **222** (colorless), **2221′** (grey), and **22′2′** (black and white). The symmetry operations associated with **1′** are colored with blue. The complete listing of SAPG and SASGs is given by Litvin (30, 31).

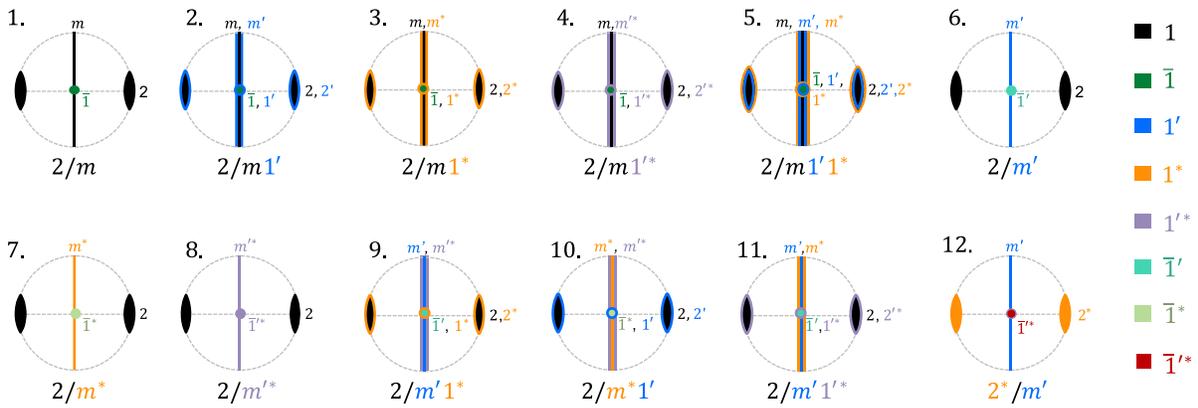

One example each of 12 types of double antisymmetry point groups (DAPGs) when 1* and 1′ are combined with the crystallographic point groups. The symmetries 1, 1′, 1*, 1′*, $\bar{1}$, $\bar{1}$′, $\bar{1}$*, $\bar{1}$′* are shown with each antisymmetry being color-coded as shown on the right. If space groups are considered, four of these symmetries involving $\bar{1}$ would not be considered antisymmetries due to the presence of translations that do not commute with $\bar{1}$.

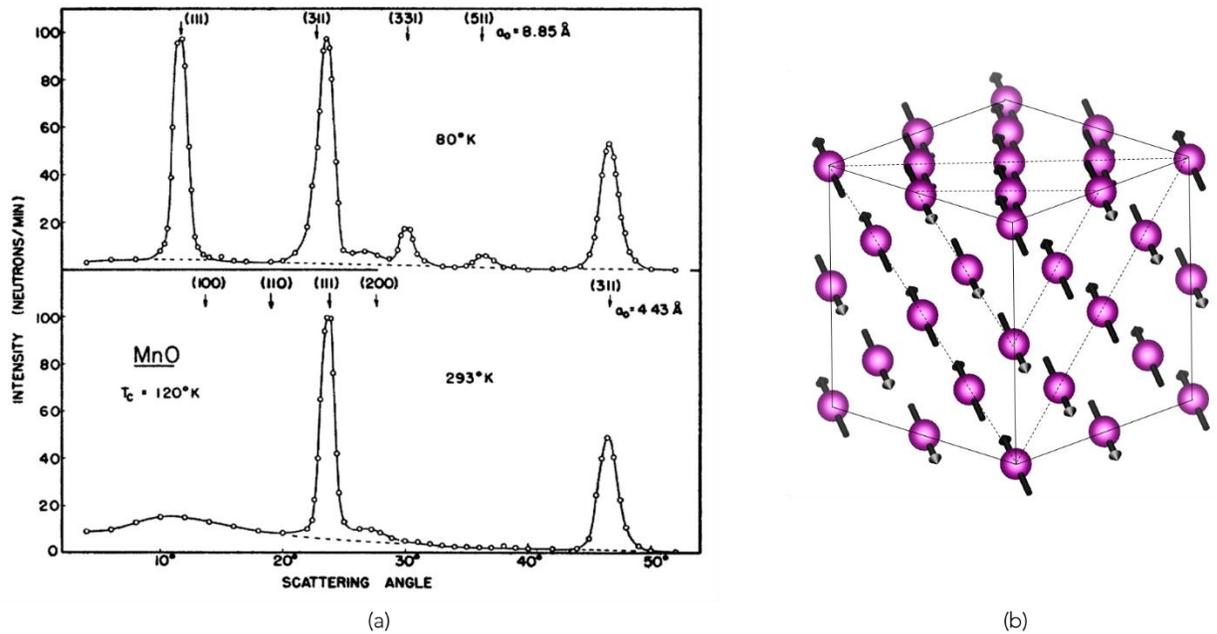

(a)

(b)

## Figure 5

The first reported (38) neutron diffraction study on MnO revealing magnetic symmetry (*a*) Unit-cell doubling and four extra antiferromagnetic reflections are observed below the Neel temperature of 120 K, since neutrons are sensitive to the spin of the scattering electrons. Figure from Shull et. al. (38). (*b*) The unit cell of MnO, shown with only the Mn atoms and the magnetic moments at each site. The dashed lines show planes within which magnetic moments are aligned. Reprinted figure with permission: C. G. Shull, W. A. Strauser, and E. O. Wollan, Physical Review, Volume 83, No. 2, 333, 1951. Copyright (2019) by the American Physical Society.

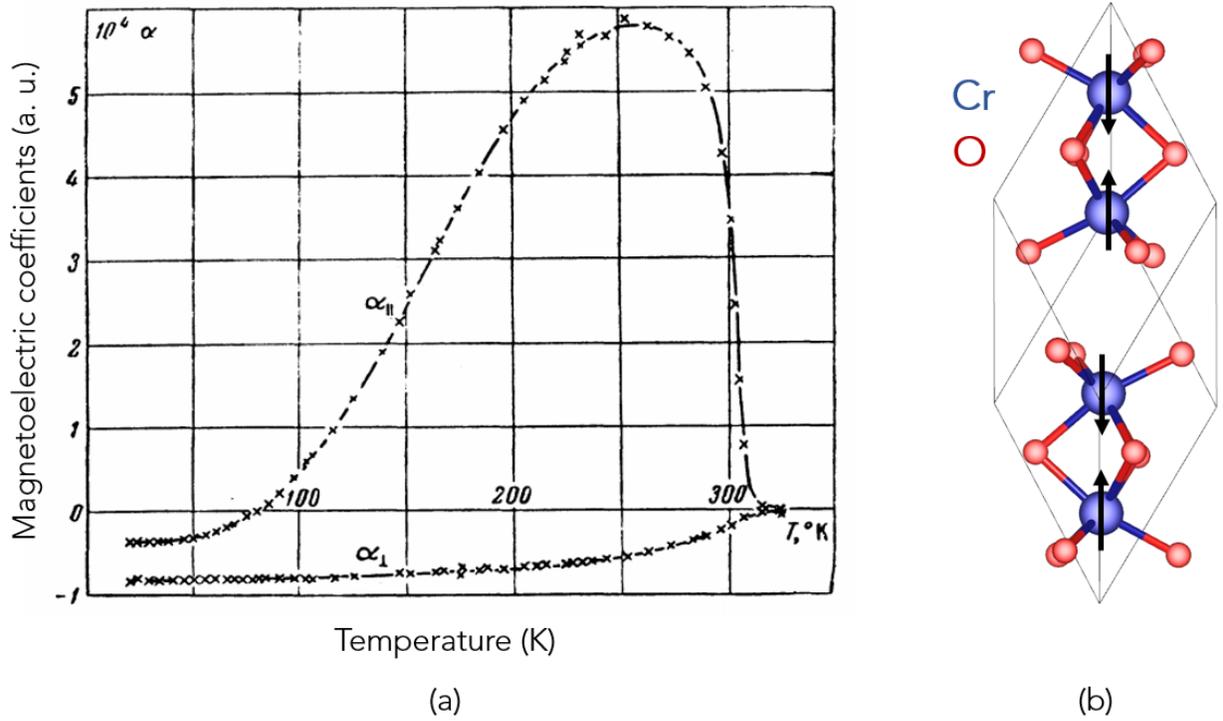

(a)            (b)

Figure 6

Magnetoelectric effect in $Cr_2O_3$. Panel (*a*) shows the first experimental discovery of the magnetoelectric, in $Cr_2O_3$. The curve $\alpha_\parallel$ corresponds to the $Q_{33}$ component in the text, and the curve $\alpha_\perp$ corresponds to the $Q_{11}$ component in the text. Figure adapted from Astrov (39). Panel (*b*) shows the crystal structure of $Cr_2O_3$, with the magnetic moments shown as black arrows. Reprinted figure with permission: JETP 13, 729 (1961) by D. N. Astrov. Copyright (2019) by the Journal of Experimental and Theoretical Physics.

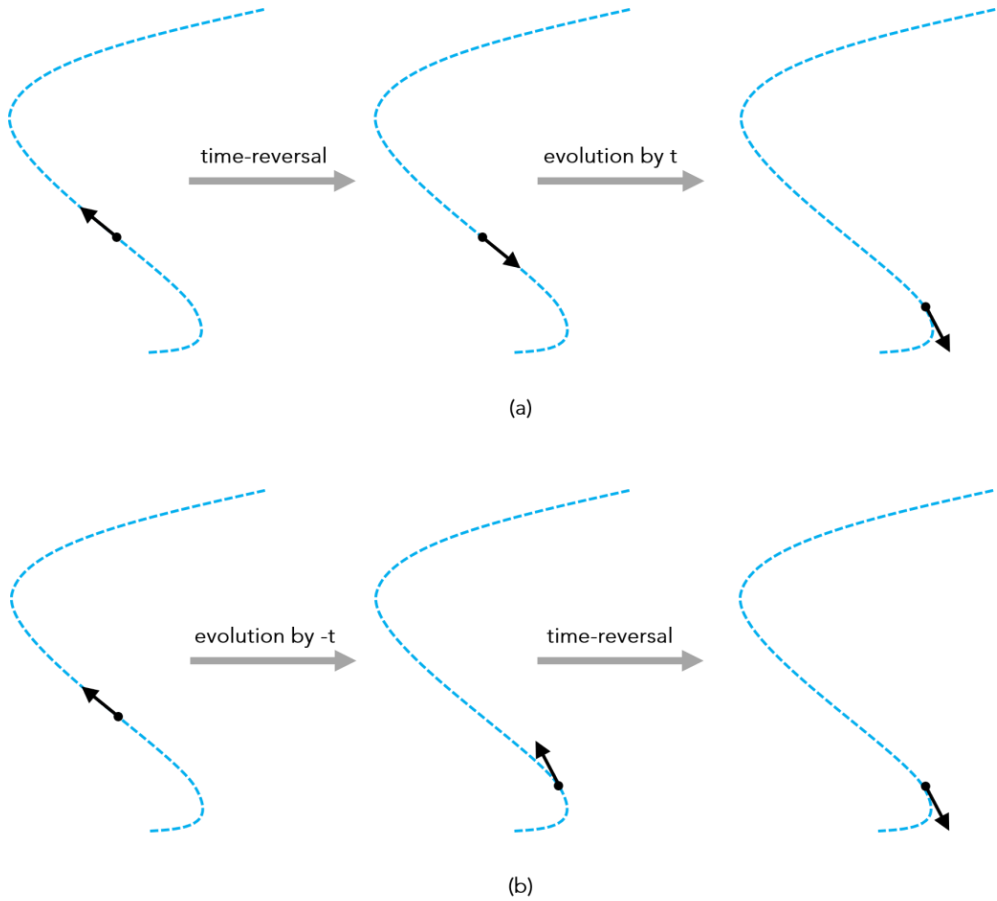

Illustration of classical time-reversal symmetry, $\mathcal{T}$. The figure shows that for a classical particle on a time-reversal symmetric trajectory, ($a$) applying the time-reversal symmetry operator first will cause the particle to reverse its momentum and retrace its trajectory, so that its new state after time $t$ is identical to ($b$) that obtained by propagating the particle by $-t$ first and then reversing its momentum. In other words, $\mathcal{T}\big(x(t), p(t)\big) = \big(x(-t), -p(-t)\big)$.

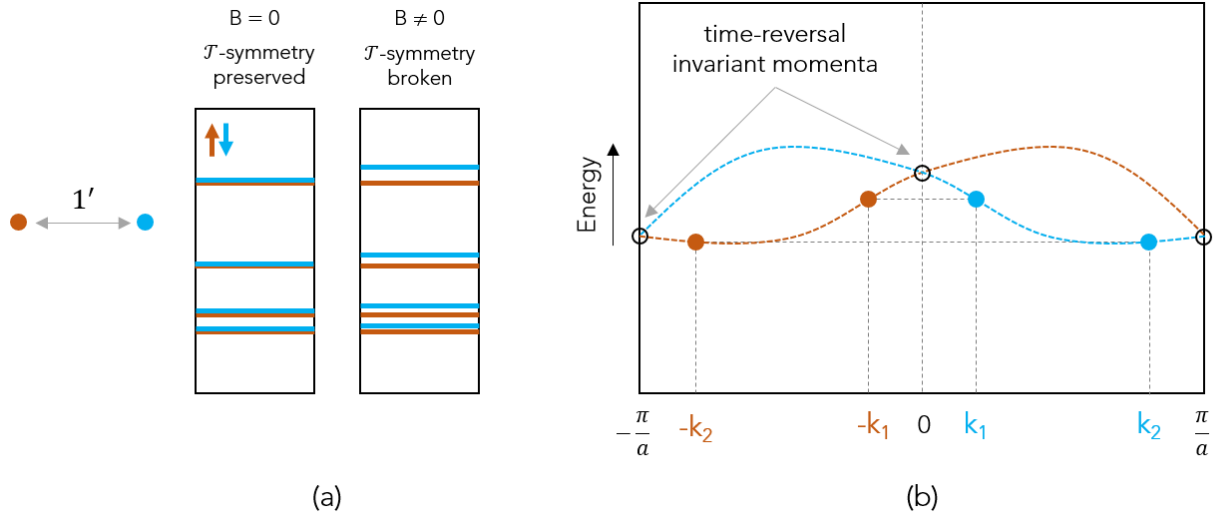

(a)

(b)

## Figure 8

Illustration of time-reversal symmetry and Kramer's degeneracy. Panel (a) shows the eigenstates of a spin-1/2 particle when time-reversal symmetry is present ($\boldsymbol{B} = 0$), and when it is broken ($\boldsymbol{B} \neq 0$). The eigenstates are necessarily two-fold degenerate in the presence of time-reversal symmetry. Panel (b) shows a band structure in the presence of time-reversal symmetry. Kramer's degeneracy implies that eigenstates at $\boldsymbol{k}$ and $\boldsymbol{-k}$ are degenerate, time-reversal symmetric pairs, so that the band structure is symmetric about $\boldsymbol{k} = 0$. Another consequence is that the eigenstates at time-reversal invariant momenta $\boldsymbol{k} = 0$, and $\boldsymbol{k} = \pm\frac{\pi}{a}$, shown with open circles, are two-fold degenerate.

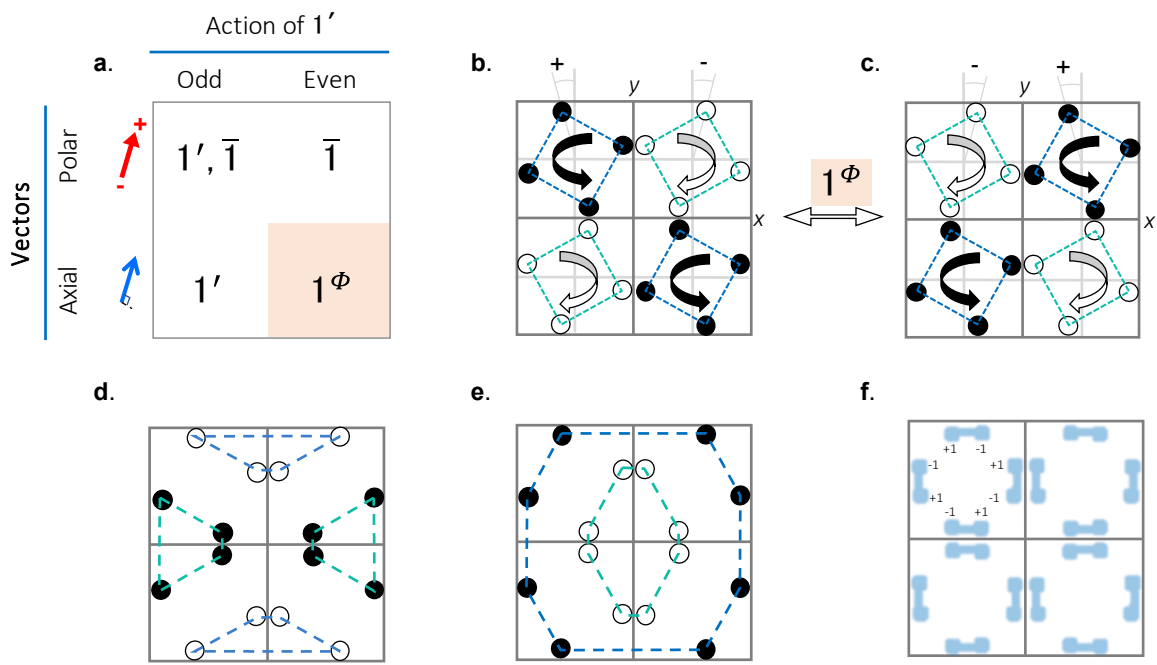

## Figure 9

Illustration of the rotation reversal antisymmetry operation $1^{\Phi}$. (*a*) If the action (even or odd) of time reversal, $1'$, on two types of vectors, polar and axial, is considered, one can place antisymmetry operations entries in each box that will reverse the corresponding vector. (*b*) and (*c*) An example illustrating the action of $1^{\Phi}$ on a collection of atoms represented by white and black circles; the atoms are all identical, but the black and white colors highlight the color symmetry. (*d*) and (*e*) illustrate a problem with the conceptual implementation of $1^{\Phi}$ – using a different selection of polyhedral units, the action of $1^{\Phi}$ will no longer be a pure rotation but a general distortion. (*f*) A general distortion can be parametrized by a dimensionless parameter, $-1 \leq \lambda \leq 1$, and reversed by a distortion reversal antisymmetry, $1^*: \lambda \rightarrow -\lambda$.

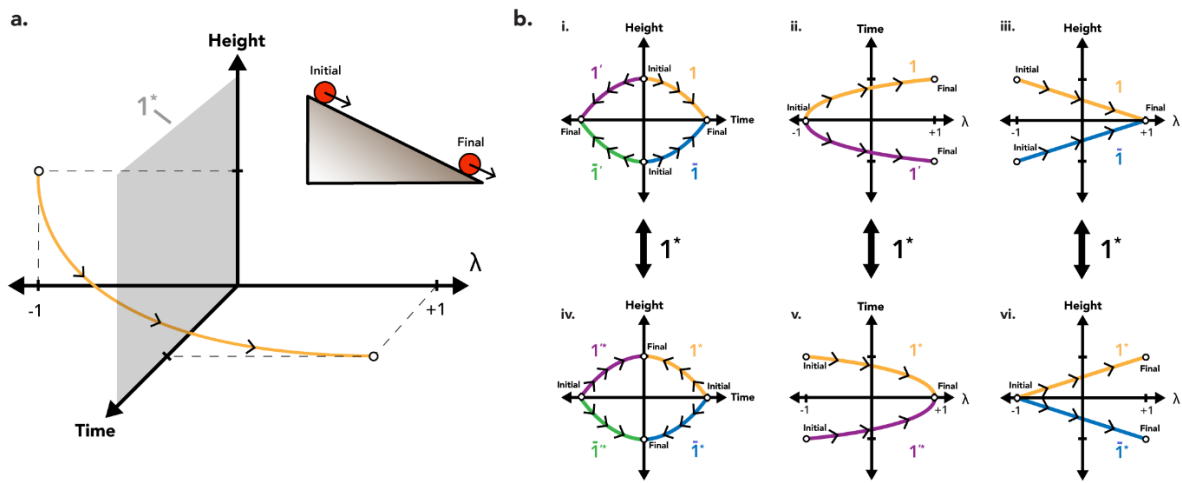

**Figure 10**

Illustration of distortion reversal antisymmetry, 1*. (*a*) Parameterization of a path taken by a ball rolling down a hill (inset) with the height being used to generate a linear reaction coordinate λ between -1 (initial) and +1 (final) states, which acts as a "time-like" coordinate. (*b*) Illustration of how both time reversal and distortion reversal operations transform the pathway taken by the ball in (*a*). The actions of elements of the group $\{1, \bar{1}, 1', 1^*, \bar{1}', \bar{1}^*, 1'^*, \bar{1}'^*\}$ is shown. On all traces, black arrows are drawn to indicate the initial and final points of the pathway as defined by the reaction coordinate λ.

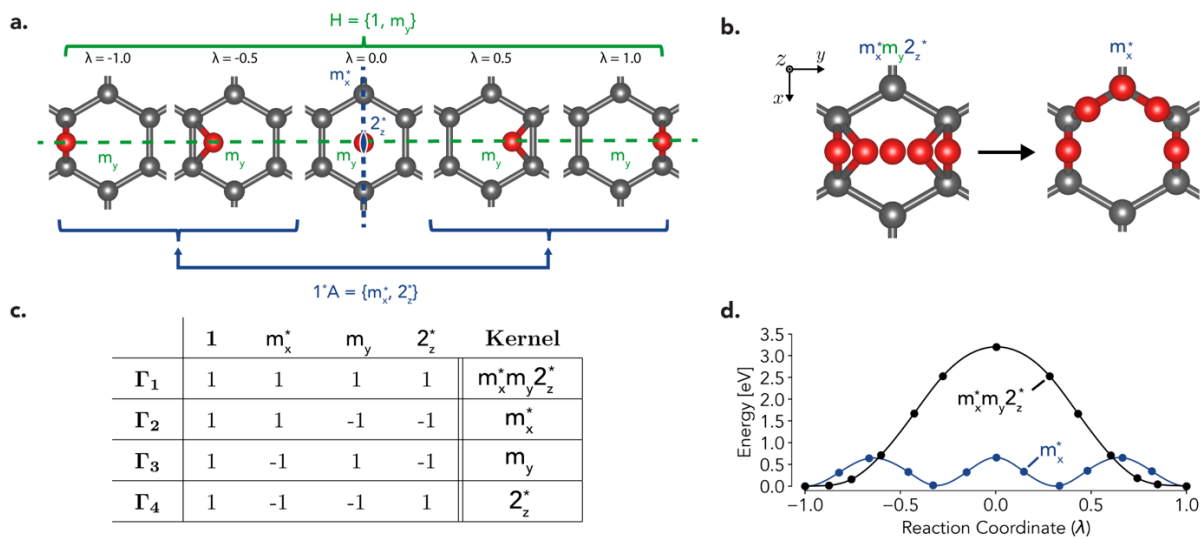

# Figure 11

The distortion symmetry group helps determine the minimum energy pathway for the diffusion of oxygen across a carbon ring in graphene. (*a*) The elements of unstarred (group H) and starred symmetry (set 1*A) elements are shown. Taking the union of both sets of elements gives the distortion group of $m^*m2^*$ for the initial path shown in (*b*) Symmetry-adapted perturbation used to lower the path symmetry to $m^*$ also shown in (*b*). (*c*) The character table of the $m^*m2^*$ shows the four irreducible representation (irreps) that define the four unique perturbations to the initial path. (*d*) Performing these perturbations in the Nudged Elastic Band (NEB) method yields the Minimum Energy Pathway (MEP) to be $m^*$. Further perturbations of the MEP by treating it as the initial path and performing the same procedure as on $m^*m2^*$ does not lower the energy barrier any further (10).

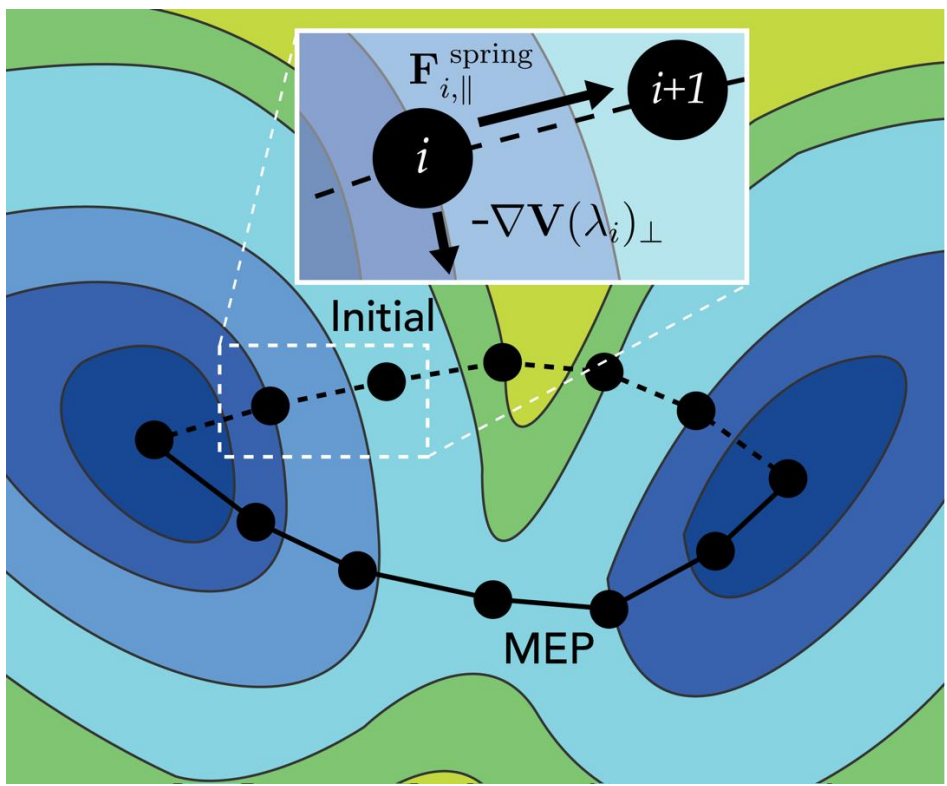

## Figure 12

Schematic illustrating the Nudged Elastic Band (NEB) method to find minimum energy pathway between two low energy valleys. The background is the potential energy(V) surface plotted as a colored contour plot with the black contour lines representing isoenergy lines. The blue end of the color spectrum represents low energy and the yellow end the high energy. Two paths are illustrated: the one with broken line is a general path, and the one with solid black line is the minimum energy path (MEP). The black filled circles represent some intermediate images (label $i$ in the inset) along the two paths, parametrized by a value of $\lambda$ between -1 (initial state) to +1 (final state) for the respective paths. Inset shows a force component parallel to the path due to the spring and the force component perpendicular to the path due to the potential energy surface.

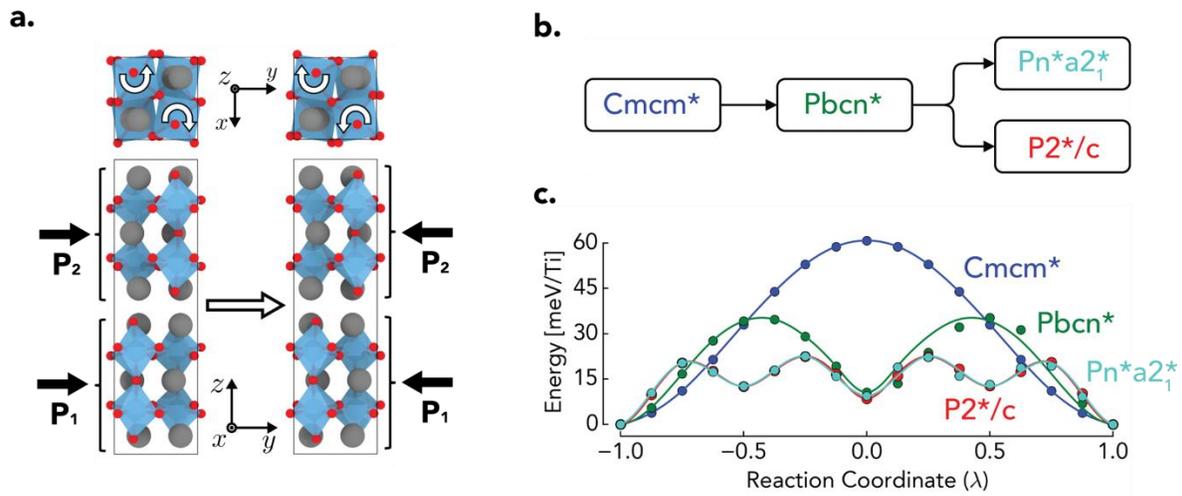

**Figure 13**

(*a*) The initial and final state of ferroelectric switching in $Ca_3Ti_2O_7$. The polarization **P** in each layer is shown with the black arrows. The Ca, Ti, and O are indicated by grey, blue and red atoms respectively. (*b*) Flowchart showing the tree of symmetry-adapted perturbations and the resulting distortion symmetry group of the path. The initial linear interpolated path between the end states has a distortion symmetry group of C*mcm*\*. (*c*) The energy profiles for the paths obtained after applying the perturbations and running the NEB algorithm. A previously reported (61) lower energy two-step path is shown in green. Two of the new lower energy four-step paths are shown in red and blue, and result from perturbing the two-step path(28). Reprinted figure with permission from <u>Jason M. Munro, Hirofumi Akamatsu, Haricharan Padmanabhan, Vincent S. Liu, Yin Shi, Long-Qing Chen, Brian K. VanLeeuwen, Ismaila Dabo, and Venkatraman Gopalan, Physical Review B, Volume 98, 085107 (2018).</u> Copyright (2019) by the American Physical Society.

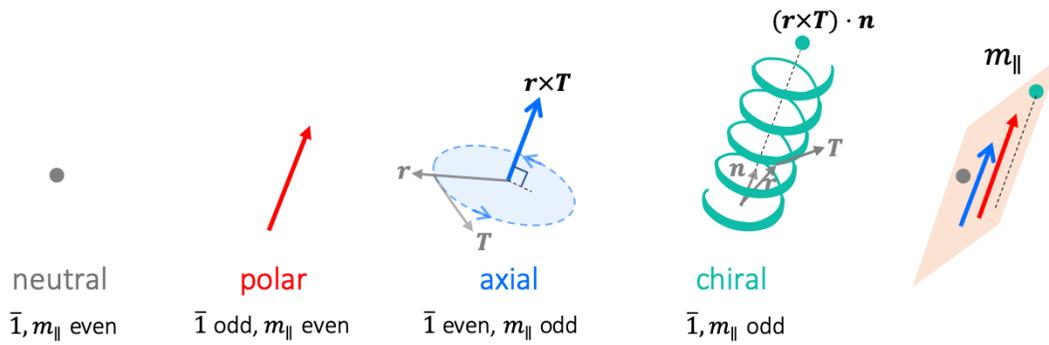

**Figure 14**

Four primary types of "vectorlike" objects in 3D (the first four from the left) classified by Hlinka (66). In 3D, these are respectively, a scalar (neutral), a polar vector (polar), an axial vector (axial), and a pseudoscalar (chiral). The transformation of these four objects under spatial inversion, $\bar{1}$ and a mirror, $m_\parallel$ defined by Hlinka is indicated below each object (even means invariant action, and odd means reverses sign under the action of the relevant symmetry operation indicated). The 2D mirror plane operation, $m_\parallel$, was defined by Hlinka as "parallel" to the scalar or to the axes of the other objects and is schematically depicted on the far right. Each of these objects can further be time-even (invariant under the action of time reversal, $1'$) or time-odd (reverse under the action of $1'$), giving rise to a total of 8 types of "vectorlike" objects.

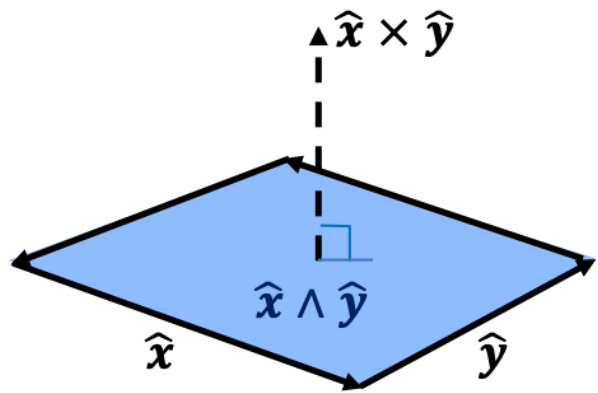

Schematic showing that the wedge product $\hat{x} \wedge \hat{y} = \hat{x}\,\hat{y}$ is a bivector whose magnitude is the unit area shown in blue with the counterclockwise circulation of vectors around its perimeter, while the cross product, $\hat{x} \times \hat{y} = \hat{z}$ is a vector whose magnitude is the unit area and which points in the direction normal to the unit area. They are Hodge duals of each other, defined as $\hat{x} \wedge \hat{y} = \hat{x}\,\hat{y}\hat{z}(\hat{x} \times \hat{y})$, where $\hat{x}\,\hat{y}\hat{z}$ is a trivector in 3D.

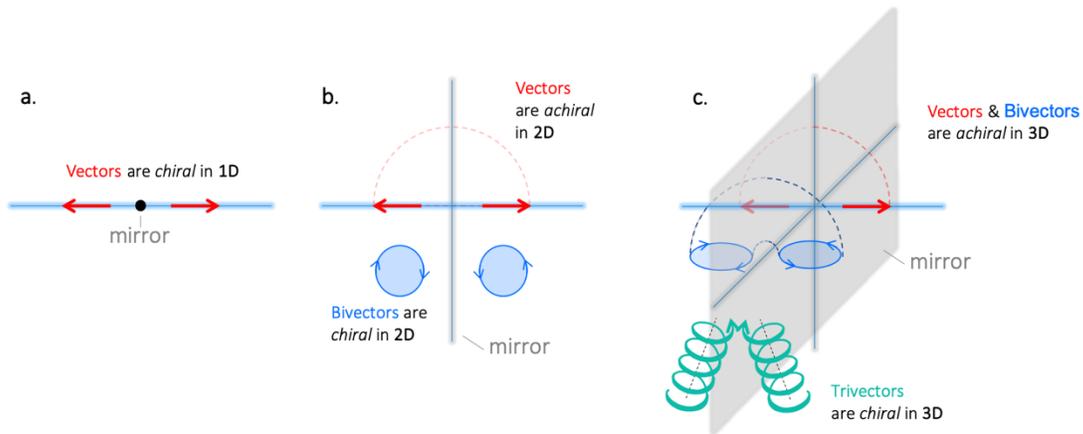

## Figure 16

Schematic showing that chirality of an object depends on the dimensionality of the ambient space it resides in. (*a*) A vector and its mirror image (red arrows) cannot be congruently overlapped in 1D, and hence it is chiral. (*b*) However, a vector and its mirror image can be overlapped congruently in 2D indicating it is achiral. However, a bivector and its mirror image (light blue circles with right-handed and left-handed circulations around their perimeters) cannot be congruently overlapped in 2D indicating they are chiral in 2D. (*c*) A vector and a bivector are both achiral in 3D as indicated by light-red and light-blue broken lines indicating the suggested trajectory for overlapping the objects and their mirror images. However, a trivector represented by a helical structure and its mirror image in light green cannot be congruently overlapped in 3D, and hence it is chiral; it will no longer be chiral in 4D and higher dimensions.

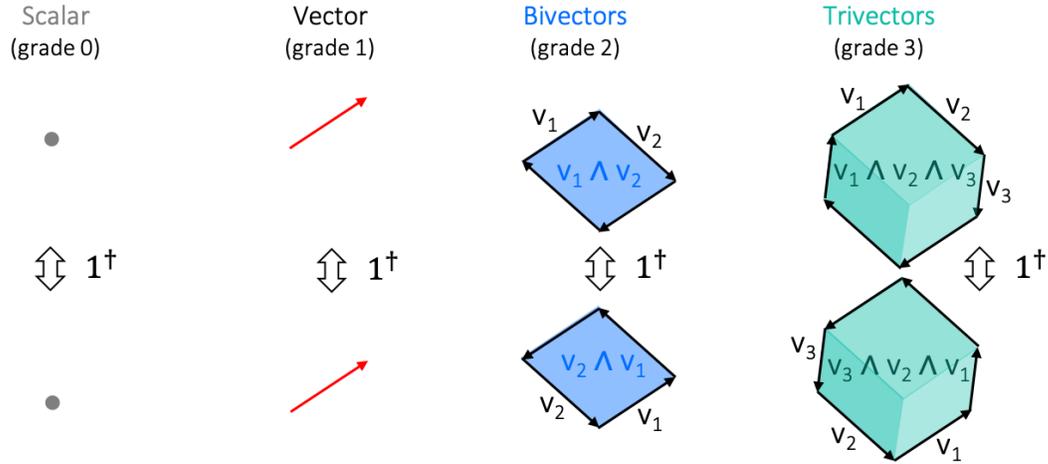

The action of wedge reversion, $1^\dagger$, on multivectors of grades 0, 1, 2, and 3. Specifically, $1^\dagger(s) = s$, $1^\dagger(\mathbf{A}) = \mathbf{A}$, and

$1^\dagger\left(\mathbf{A^{(1)}} \wedge \mathbf{A^{(2)}} \wedge \mathbf{A^{(3)}} \ldots \wedge \mathbf{A^{(n-1)}} \wedge \mathbf{A^{(n)}}\right) = \mathbf{A^{(n)}} \wedge \mathbf{A^{(n-1)}} \ldots \wedge \mathbf{A^{(3)}} \wedge \mathbf{A^{(2)}} \wedge \mathbf{A^{(1)}}$, where $s$ is a scalar, $\mathbf{A}$ is a vector, and $\mathbf{A^{(i)}}$ ($i$ is vector index= 1, 2, 3…. $n$) are $n$ linearly independent vectors. Bivectors and Trivectors reverse under the action of $1^\dagger$, while scalars and vectors do not. Note that unlike Figure 13, no reference is made to axiality (cross products) or the chirality (which requires one to consider the dimensionality of the ambient space the multivector resides in); these descriptions are dropped. Wedge reversion is well-defined for an arbitrary grade multivector residing in an arbitrary dimension of the ambient space.

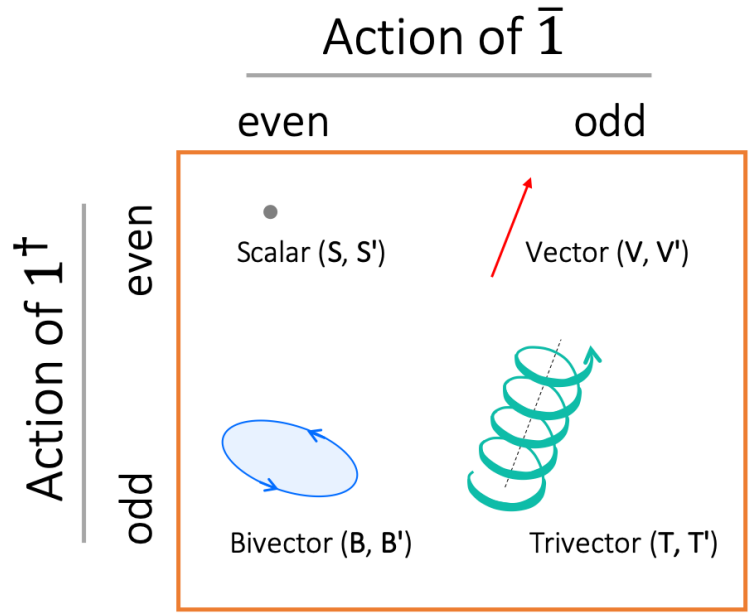

## Figure 18

Eight types of multivectors in 3D. They are labeled S', V', B', and T' for time-even and S, V, B, and T for time-odd multivectors. The axial vector quantity, $\boldsymbol{r} \times \boldsymbol{T}$ from Figure 13 would instead be treated as a bivector quantity, $\boldsymbol{r} \wedge \boldsymbol{T}$ in the above classification. Similarly, the chiral quantity $(\boldsymbol{r} \times \boldsymbol{T}) \cdot \boldsymbol{n}$ from Figure 13 would instead be treated as a trivector quantity, $\boldsymbol{r} \wedge \boldsymbol{T} \wedge \boldsymbol{n}$ in the above classification. In general, the classification proceeds as multivectors of grades $4g$, $4g+1$, $4g+2$, and $4g+3$, $g$=0, 1, 2, 3..., respectively being either S', V', B', and T' for time-even, or S, V, B, and T for time-odd multivectors (see Table 1). The case of $g$=0 is depicted above.

**Table 1: Antisymmetry-based classification of 8 principal types of multivectors.**

| # | Label | Action of $\overline{1}$ | $1'$ | $1^{\dagger}$ | Type | Grade | Examples |
|---|-------|---|---|---|------|-------|----------|
| 1) | S' | $e$ | $e$ | $e$ | $t$-even *centric-acirculant* | $4g$ | $t^2$, $(\boldsymbol{r} \cdot \boldsymbol{P})$, $((\boldsymbol{r} \times \boldsymbol{P}) \cdot \boldsymbol{n})$ |
| 2) | V' | $o$ | $e$ | $e$ | $t$-even *acentric-acirculant* | $4g+1$ | $\boldsymbol{r}, \boldsymbol{P}, \boldsymbol{E}$ |
| 3) | B' | $e$ | $e$ | $o$ | $t$-even *centric-circulant* | $4g+2$ | $(\nabla \wedge \boldsymbol{P})$, $(\boldsymbol{r} \wedge \boldsymbol{P})$ |
| 4) | T' | $o$ | $e$ | $o$ | $t$-even *acentric-circulant* | $4g+3$ | (B')$\wedge \boldsymbol{n}$, *e.g.* $\boldsymbol{r} \wedge \boldsymbol{P} \wedge \boldsymbol{n}$ |
| 5) | S | $e$ | $o$ | $e$ | $t$-odd *centric-acirculant* | $4g$ | $t$, $(\boldsymbol{r} \cdot \boldsymbol{p})$, $((\boldsymbol{r} \times \boldsymbol{J}) \cdot \boldsymbol{n})$ |
| 6) | V | $o$ | $o$ | $e$ | $t$-odd *acentric-acirculant* | $4g+1$ | $\boldsymbol{v}, \boldsymbol{J}, \boldsymbol{p}, (\nabla \times \boldsymbol{S})$, $(\boldsymbol{E} \times \boldsymbol{H})$ |
| 7) | B | $e$ | $o$ | $o$ | $t$-odd *centric-circulant* | $4g+2$ | $*\boldsymbol{M}$, $*\boldsymbol{S}$, $*\boldsymbol{B}$, $\boldsymbol{r} \wedge \boldsymbol{J}$ |
| 8) | T | $o$ | $o$ | $o$ | $t$-odd *acentric-circulant* | $4g+3$ | B$\wedge \boldsymbol{n}$, *e.g.* $\boldsymbol{r} \wedge \boldsymbol{J} \wedge \boldsymbol{n}$ |

Labels for 8 different multivectors types (Column 1) is introduced in Column 2. Columns 3-5 presents the action of three antisymmetries, namely, $\overline{1}, 1'$, and $1^{\dagger}$ on these multivectors as either even ($e$ for invariant), or odd ($o$, odd for sign reversal). Column 6 presents the type of multivector, and column 7 gives the grade of the multivector, where $g$=0, 1, 2, 3…., a whole number. Column 8 presents some examples of multivectors in 3D. Conventional polar vectors are presented in ***bold italics***. Multivector types (column 2) are presented as CAPITAL letters without italics or without bold. Bold italics are reserved for vectors, such as $\boldsymbol{P}$, which is time-even polarization vector measured in units of C/m$^2$. The vectors $\boldsymbol{r}$ and $\boldsymbol{n}$ are two linearly independent vectors. Other definitions are as follows: $\boldsymbol{E}$ is electric field, $\boldsymbol{v}$ is velocity, $\boldsymbol{J}$ is current density, $\boldsymbol{p}$ is momentum, $\boldsymbol{H}$ is magnetic field, $\boldsymbol{B}$ is Magnetic induction, $\boldsymbol{S}$ is spin, $t$ is time. The * as in $*\boldsymbol{B} = \hat{\boldsymbol{x}}\hat{\boldsymbol{y}}\hat{\boldsymbol{z}}\,\boldsymbol{B}$ in 3D, and similarly for others. For $n$ dimensions, $\hat{\boldsymbol{x}}\hat{\boldsymbol{y}}\hat{\boldsymbol{z}}$ is replaced by the geometric product of the $n$ basis vectors spanning that dimension. The * operation is the Hodge dual operation in Clifford Algebra. A wedge product, B$\wedge \boldsymbol{n}$, is between a position vector $\boldsymbol{n}$, and any bivector of type B